\documentclass[10pt, twocolumn]{article}
\usepackage{times}
\usepackage{epsfig}
\usepackage{graphicx}
\usepackage{amsmath}
\usepackage{amssymb}
\usepackage{enumitem}
\usepackage{soul}
\usepackage{float}

\PassOptionsToPackage{hyphens}{url}
\usepackage[colorinlistoftodos]{todonotes}

\usepackage{comment}
\usepackage{verbatim,epsfig,times,subfigure}
\usepackage{color}
\usepackage[export]{adjustbox}  
\usepackage[T1]{fontenc}

\usepackage{booktabs}

\usepackage[breaklinks=true,bookmarks=false]{hyperref}

\def\cvprPaperID{****} 

\usepackage[ top = 1.50cm, bottom = 2.50cm, left = 2.00cm, right = 1.50cm]{geometry}

\begin{document}

\title{Synthetic Control, Synthetic Interventions, and COVID-19 spread: Exploring the impact of lockdown measures and herd immunity}

\author{Niloofar Bayat\\
Columbia University\\
{\tt\small niloofar.bayat@columbia.edu}
\and
Cody Morrin\\
Columbia University\\
{\tt\small cody.morrin@columbia.edu}
\and
Yuheng Wang\\
Columbia University\\
{\tt\small yw3474@columbia.edu}
\and
Vishal Misra\\
Columbia University\\
{\tt\small vishal.misra@columbia.edu}
}

\maketitle

\begin{abstract}
The synthetic control method is an empirical methodology for causal inference using observational data. By observing the spread of COVID-19 throughout the world, we analyze the data on the number of deaths and cases in different regions using the power of prediction, counterfactual analysis, and synthetic interventions of the synthetic control and its extensions. We observe that the number of deaths and cases in different regions would have been much smaller had the lockdowns been imposed earlier and had the re-openings been done later, especially among indoor bars and restaurants. We also analyze the speculated impact of herd immunity on the spread given the population of each region and show that lockdown policies have a very strong impact on the spread regardless of the level of prior infections.

Our most up-to-date code, model, and data can be found on github: \url{https://github.com/niloofarbayat/COVID19-synthetic-control-analysis} 
\end{abstract}

\section{Introduction}\label{sec:intro}

In December 2019, an outbreak of an infectious disease was identified in China, caused severe acute respiratory syndrome coronavirus 2 (SARS-CoV-2), a.k.a. COVID-19. In March 2020, the World Health Organization (WHO) declared the COVID-19 outbreak a pandemic \cite{Mayooutbreak}. While the first case was traced back to November 17, 2019, \cite{MafirstCOVID}, by July 9, 2020, more than $550,000$ deaths have been reported globally from almost $12$ million reported cases \cite{CSSEJHU}.

During the spread of COVID-19, different regions took different measures in controlling and containing it. In some, the lockdowns were very strict and started early on, and mask mandates were strict, whereas in some others, they were more flexible, and the lockdowns started later or were removed quickly during the spread. Besides the policies, the dominant lifestyle, population, central air conditioning, etc. may impact the spread of the virus differently in different regions. Therefore, the questions of what are the best policies in each area to slow down the pandemic and how to assess the impact of individual policies remain controversial.

\begin{figure}[t]
\centering
\subfigure[global cases]{\includegraphics[width=1.65in, angle=0]{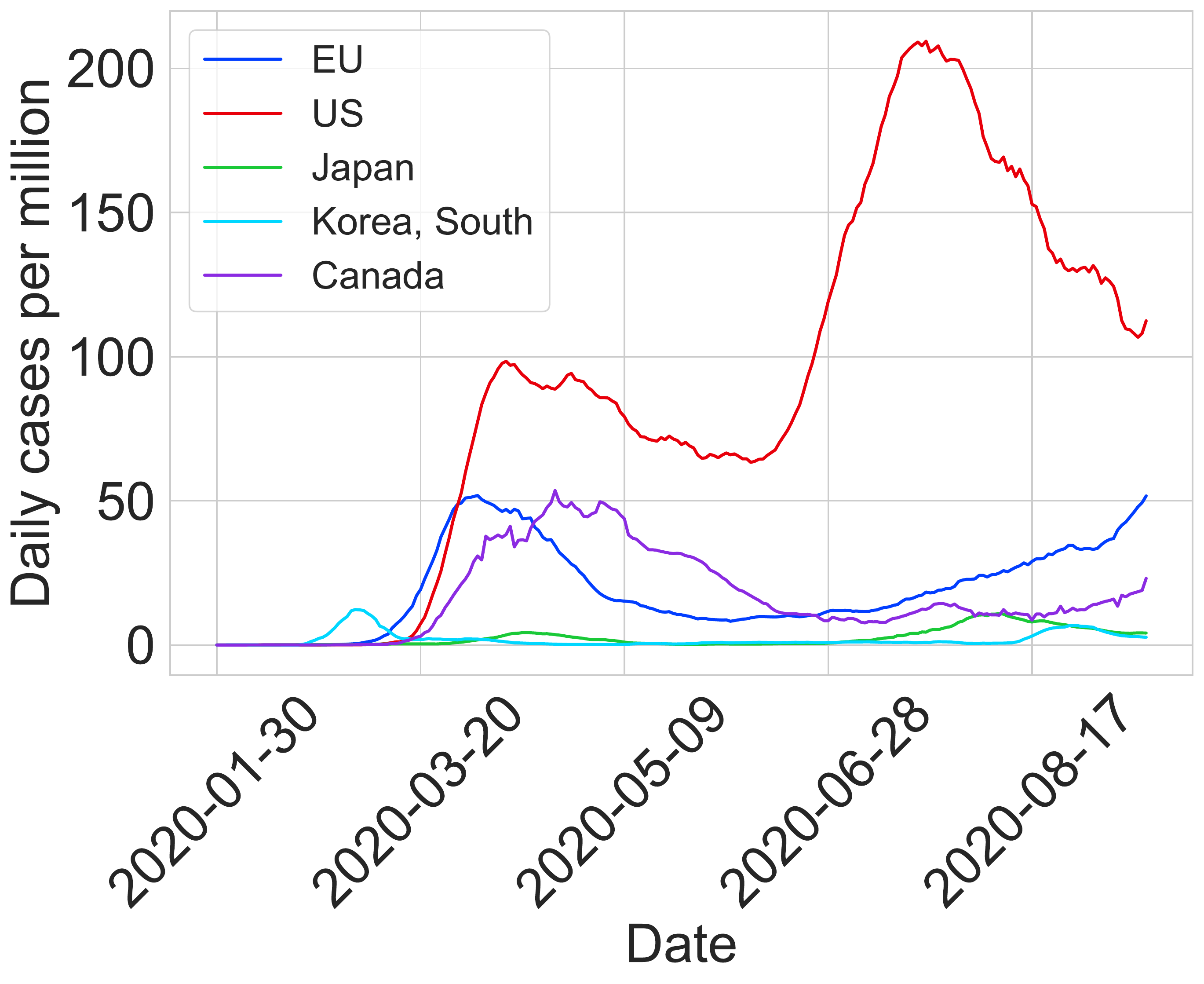}}
\subfigure[US cases]{\includegraphics[width=1.65in, angle=0]{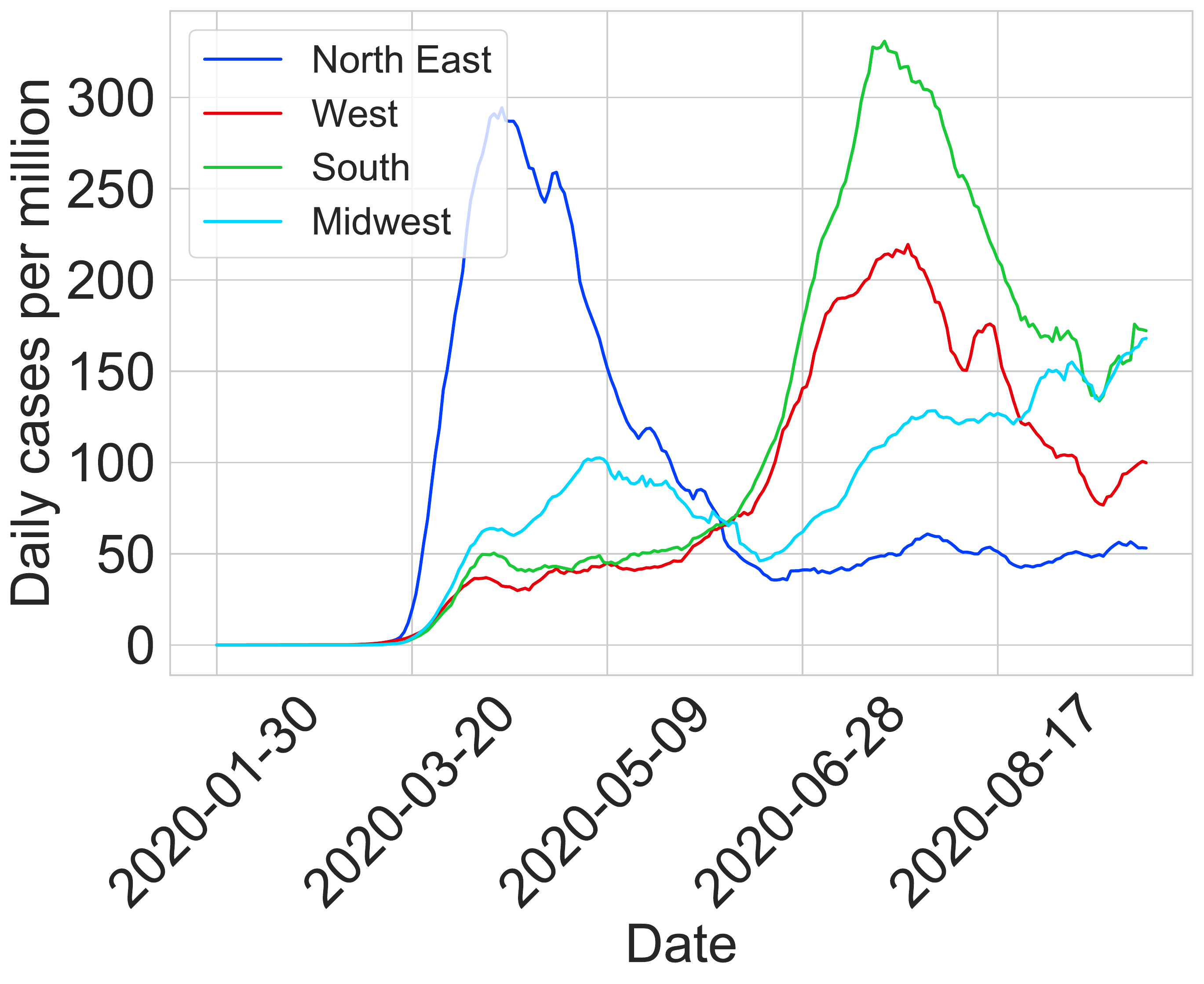}}
\caption{COVID-19 cases in a) different regions in the Globe b) different regions in the U.S.}
\label{fig:global_us_cases}
\end{figure}

\begin{figure}[t]
\centering
\subfigure[Alabama cases/deaths]{\includegraphics[width=1.65in, angle=0]{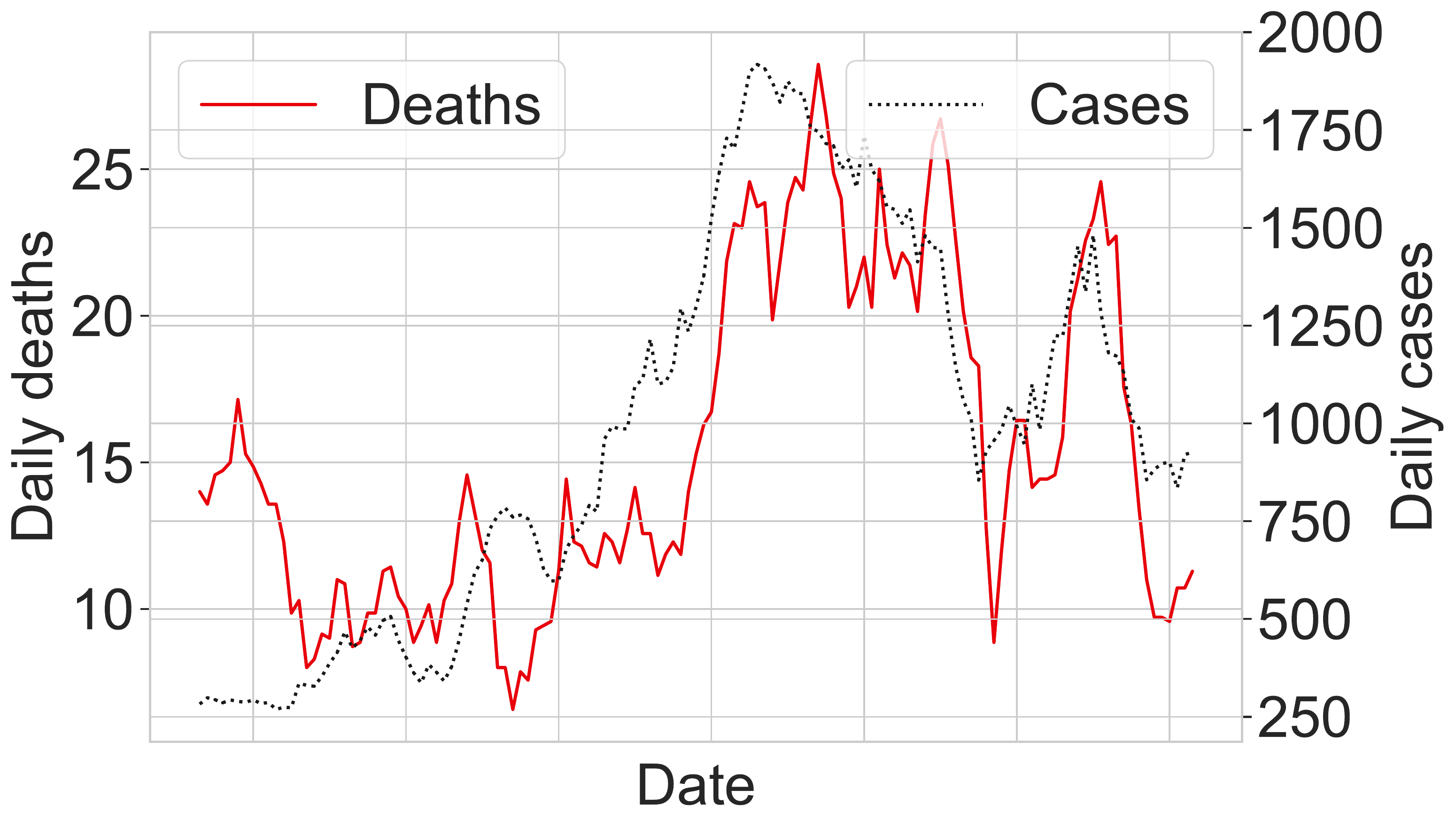}}
\subfigure[California cases/deaths]{\includegraphics[width=1.65in, angle=0]{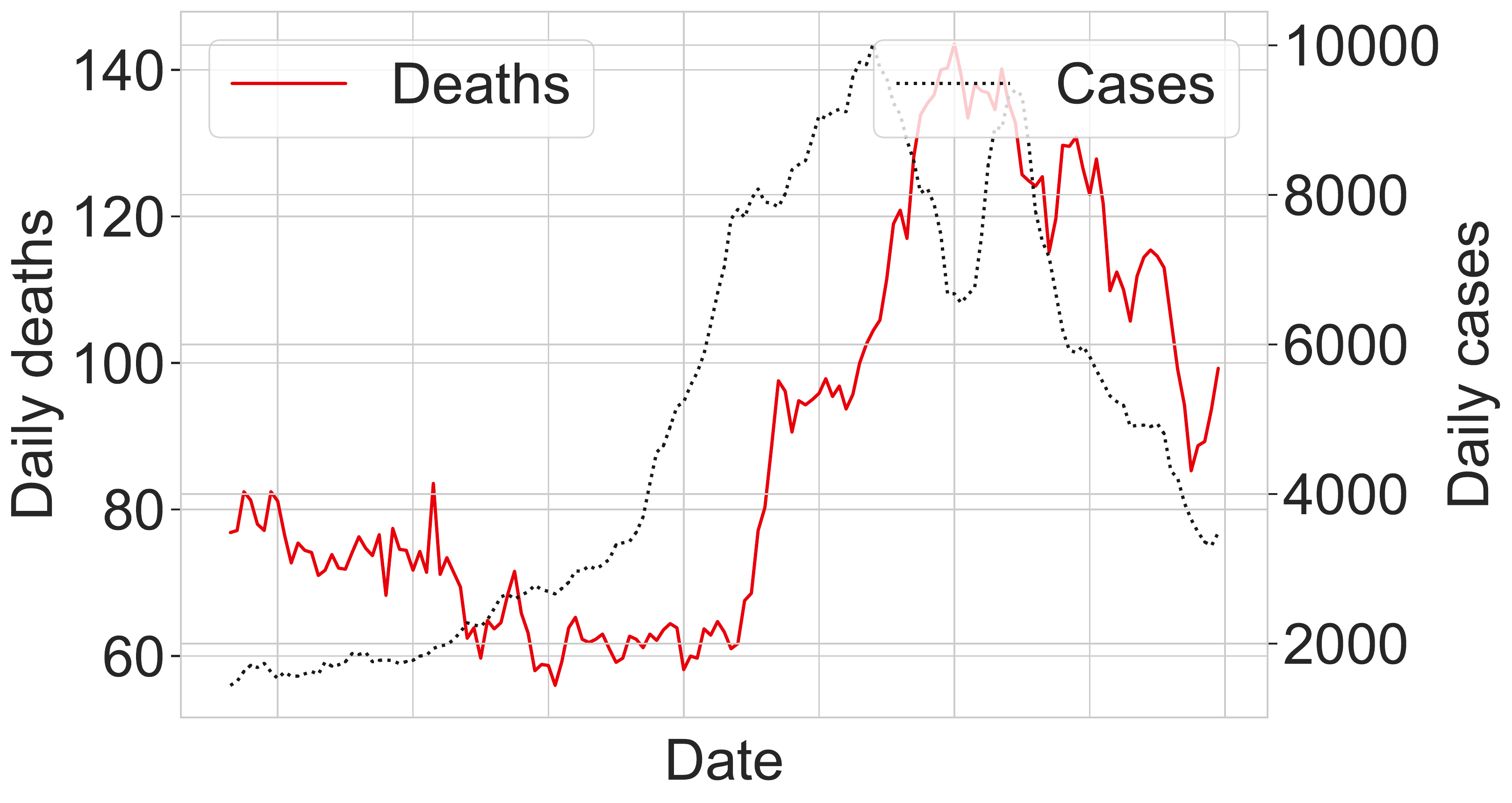}}
\subfigure[Kansas cases/deaths]{\includegraphics[width=1.65in, angle=0]{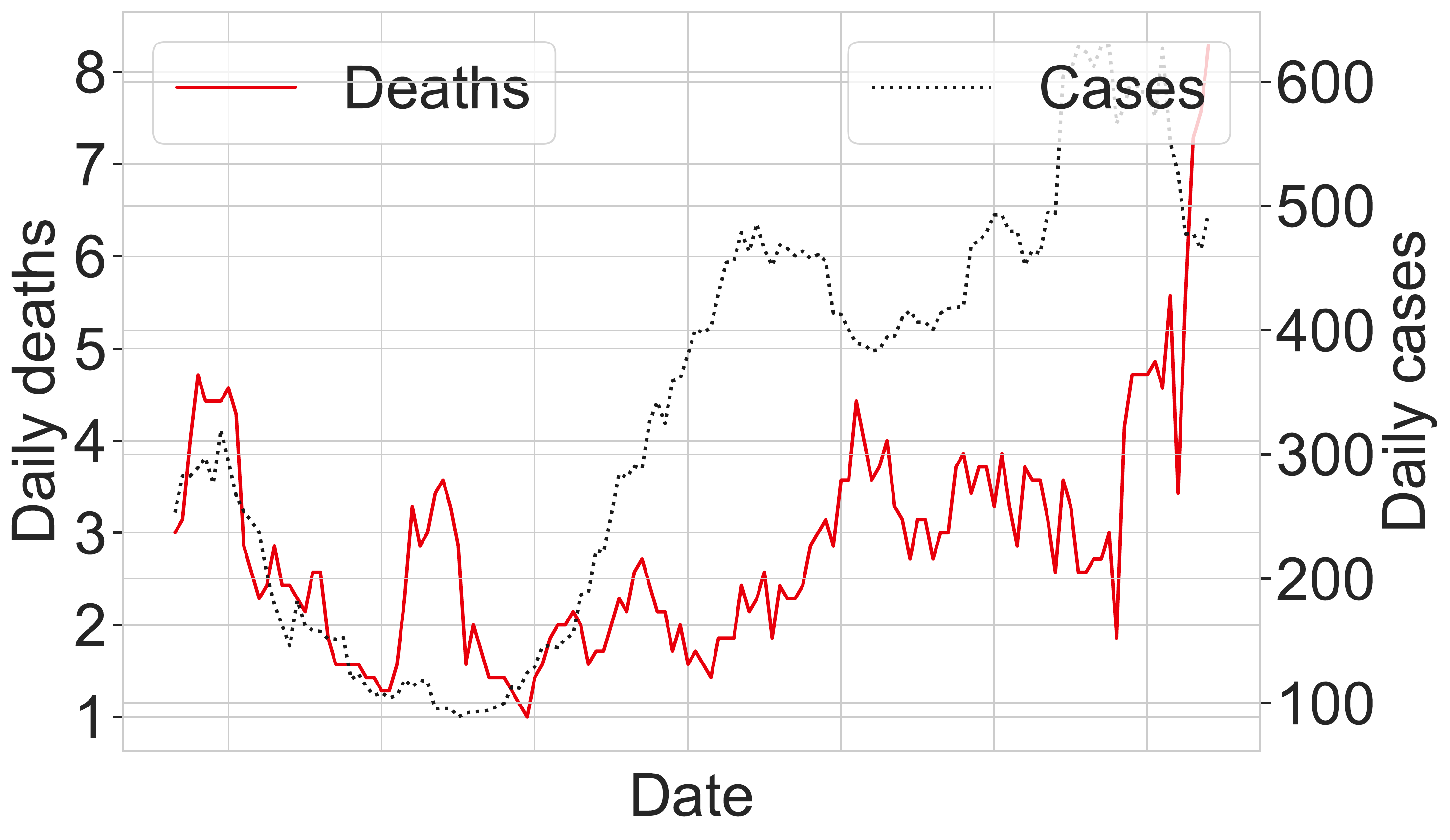}}
\subfigure[Iowa cases/deaths]{\includegraphics[width=1.65in, angle=0]{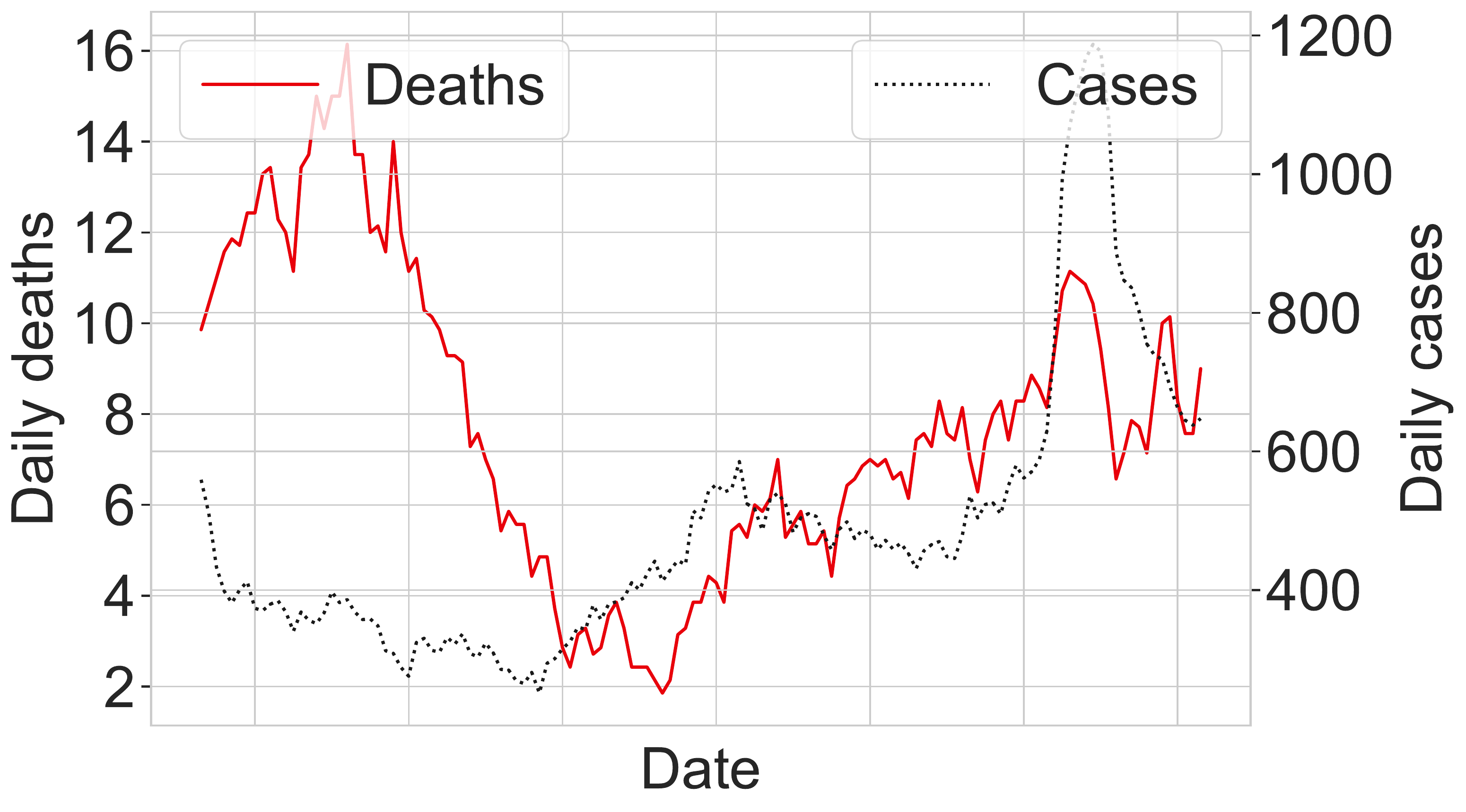}}
\caption{COVID-19 cases and deaths in four different states of the U.S.}
\label{fig:states_cases_deaths}
\end{figure}

The classical epidemiological models may fail to predict the exact pattern of COVID-19 spread. The spread of this pandemic has shown completely different behavior in different regions, both in the number of cases and in the fatality rate. Figure \ref{fig:global_us_cases} depicts the the number of cases over time in five different regions of the globe (a), and four different regions of the U.S. \cite{US_regions} (b)\footnote{Note that in all figures of this paper which depict the number of daily cases/deaths, the moving average over $7$ days is presented rather than the absolute value, to minimize the effect of potential noise in the existing reports.}. Furthermore, Figure \ref{fig:states_cases_deaths} depicts four different examples of how the spread and fatality rate vary in different states of the U.S. From these examples, this is apparent that we need a data-driven, non-parametric way to look at things, and classical epidemiological models fail to do so. Therefore, we have introduced our synthetic control model to analyze the virus spread in different regions with different patterns.

Abadie and Gardeazabal \cite{abadie2003economic} and Abadie et al. \cite{abadie2010synthetic} pioneered the synthetic control method to address the problem of measuring the impact of a new regulation or a change in a region, and comparing it to the case that those changes had not happened. The idea of synthetic control stems from the classical A/B testing, where two versions of a variable are compared in the otherwise identical environment \cite{young2014improving}. A variation of that would be where one of the variables is a placebo. In that case, the experimental units are called ``treatment'' and ``control'' groups, and the variable changes, i.e., treatments, are applied to the treatment group \cite{chaplin2006placebo}. Abadie et al. suggested that since in some problems we cannot have an actual control group, we construct a ``synthetic'' control group for the treatment unit, using the untreated units as the donor pool.

The synthetic control is a statistical method used to evaluate the impact of an intervention in observational data by performing comparative case studies. In comparative case studies, researchers estimate the evolution of aggregate outcomes
(such as mortality rates, average income, crime rates, etc.) for a
unit affected by a particular occurrence of the event or intervention and compare it to the evolution of the same aggregates estimated for some control group of unaffected units \cite{abadie2010synthetic}.

In the analysis of COVID-19 data on the number of cases and deaths, there might be some missing data that may cause an additional noise we need to account for. COVID-19 tests may not have been done extensively in numerous regions, especially in the early stages of the spread. Furthermore, in most places, the death data only accounts for hospitalization cases. Since the classical synthetic control method may not perform well given some randomly missing data and the high variance noise, we use robust synthetic control, which is a generalization of the late \cite{Amjad}. The robust synthetic control method performs ``de-noising'' estimations that are theoretically sound and computationally efficient. Additionally, we employ the recently developed extension of the synthetic control method known as synthetic interventions \cite{SyntheticInterventions}, to predict the impact of different interventions in different regions.

Our main contributions and findings are listed below:
\begin{itemize}
 \item We build a counterfactual model to measure to what extent lockdown and reopening dates impact the spread.
 \item We demonstrate via a counterfactual model the impact of imposing the lockdown earlier in New York, with our estimates predicting an upwards of 80\% reduction in the number of deaths.
 \item We analyze the impact of bars and indoor dine-in in causing the spread, where the former has a stronger impact on the spread than the latter. We also analyze the impact of the speed of reopening in causing the second wave observed in the US. 
 \item We explore the question of "herd immunity" and whether the infection level has a stronger impact on the further spread of the virus compared to social distancing measures. Our results indicate that social distancing measures, at current infection levels in the US, have a \emph{measurable} impact on the spread. 
\end{itemize}

The rest of the paper is organized as follows: Section \ref{sec:related-work} provides a brief review of the existing related work. In Section \ref{sec:background} we provide a background on synthetic control. Section \ref{sec:methods} introduces our methodology to apply robust synthetic control to the problem of COVID-19 spread and regulations impact, and Section \ref{sec:results} presents our findings. 
Finally, the paper is concluded in Section \ref{sec:conclusion}.

\section{Related Work}\label{sec:related-work}
Abadie \textit{et al}. first introduced synthetic control to measure the impact of political instability on economic prosperity \cite{abadie2003economic}, in which They investigate the economic impact of conflict, using the terrorist conflict in the Basque Country as a case study. They used the combination of other regions in Spain to build a ``synthetic'' control region which resembles economic characteristics of the Basque Country before the outset of terrorism attack. After that, the method has been widely applied in econometric of policy evaluations including studying the effects of laws \cite{donohue2019right}, legalized prostitution \cite{cunningham2018decriminalizing} and immigration policy \cite{bohn2014did}. 

The method has also been utilized outside economics: in the social sciences, biomedical disciplines, engineering, etc. For example, Kreif \textit{et al}. used the synthetic control approach to a setting of evaluation of P4P (pay-for-performance) health policy where there were multiple treated units \cite{kreif2016examination}. Opposing to general belief by that time, they showed that P4P significantly increased the mortality of non‐incentivised conditions .On the other hand, Doris \textit{et al} investigates if natural disasters help or hurt politician's electoral fortunes \cite{heersink2017disasters}, in which they study on a case of catastrophic flooding in the American South in 1927, and also showed that use of synthetic control methods and suggest that—even if voters distinguish between low- and high-quality responses—the aggregate effect of this disaster remains broadly negative.

\section{Background}\label{sec:background}
One of the technical challenges in modeling the impact of any interventions is accurately modeling the spread of the infection. This depends on many factors including the characteristics of the disease and its spread, and also prevailing societal norms. There is a vast literature on epidemic modeling including the classical SIR and SEIR models \cite{kermack27} that consider a fluid approximation for the spread of the infection. While these give a reasonable approximation in many settings, the model predictions are extremely sensitive to the parameters. Since these are estimated from noisy time series data, these are inherently subject to statistical estimation error leading to a higher variance in model predictions. Moreover, these fluid models are largely {\em homogeneous} require several simplifying assumptions that limit the ability to modeling many aspects or features specific to a particular region. These pose a significant hurdle in reasonably modeling the impact of various interventions for policy decisions based on these fluid approximations.

The goal of this work is to develop approaches to model the impact of pandemic interventions that overcome the limitations of standard approaches. In particular, our first goal is to develop a non-parametric approach based on synthetic control to study the impact of interventions. Most existing methods in epidemiology, including the SEIR model, focus on the dynamics of the size of the population within each category. These models often involve some differential equations for the dynamics of population size and relationship among different categories, which are mostly simple and straightforward. These models remain powerful tools to understand the transmission dynamics of the coronavirus, however, the model assumptions can be violated due to complicated circumstances, especially when various non-pharmaceutical interventions are implemented day by day.

These concerns motivate us to develop more flexible and non-parametric approaches to understand the dynamics of the spread of the coronavirus. Another concern, especially when trying to understand the intervention effects, is possible bias from naive comparisons as discussed a lot in the causal inference literature. Specifically, the usual comparison between pre-intervention and post-intervention periods may suffer from biases due to unobserved time trends. For example, the reduction in transmission rates of the coronavirus may be due to weather change instead of the new social distancing policy. Also, the usual comparison between intervened city/country and non-intervened one may suffer from selection bias. For example, the city which implemented strict policies (such as stay-at-home) may suffer from a more severe attack of the coronavirus, and simply comparing cities with and without stay-at-home policy may underestimate the policy effects.

Due to the concerns listed above, we propose to use synthetic control to infer the causal effects of the interventions of interest, based on which we hope to provide useful policy suggestions to mitigate the spread of the coronavirus. The synthetic control methods try to find a linear combination of non-intervened regions as an approximation (i.e., a synthetic control) of the intervened region, by matching pre-intervention characteristics such as population size and socioeconomic status. We then compare the post-intervention period of the intervened region and the synthetic control to estimate the intervention effects. 

\subsection{Synthetic Control and Proposition 99}\label{sec:prop99}

A canonical application of the synthetic controls method is the analysis of the policy of Prop99 on cigarette sales. In November 1988, California voters approved the California Tobacco Tax and Health Protection Act of 1988, also known as Prop 99 \cite{Prop99}. This initiative increased the state cigarette tax by $25$ cents per pack and added an equivalent amount on other tobacco products. Aside from making tobacco products more expensive to the customers, the revenue raised was used for various environmental and health care programs, talks in schools and colleges on the harms of tobacco, and anti-tobacco advertisements and research. Eventually, after this proposition, smoking reduced in the state of California. But there is an important question regarding the effectiveness of this policy: was the reduction of smoking because of Prop 99 or did society move on from smoking? Since the nationwide data showed the reduction of smoking on average as well, the question is how to assess the role Prop 99 played in that reduction in California.

One way to answer this question in an ideal world would be if we had two California states, one with Prop 99, and one without it. But unfortunately, this cannot happen. To answer this question, Abadie et al. \cite{abadie2010synthetic} proposed the synthetic control model to represent California as a linear combination of the other $49$ states of the U.S.

Suppose we have a smoking matrix for the U.S. Let's assume the columns of this matrix are the time series of per-capita smoking, and each row represents a different state. Although smoking habits are not the same in different states, there are some correlations among per-capita smoking due to the level of awareness to the smoking, the state of economy, etc. In synthetic control, we take the smoking data of California for a period of time right before Prop 99 was implemented (1988). For that given period, we try to approximate the time series of California as a linear combination of the other states. 
The high-level algebraic idea is that the smoking matrix is approximately low-ranked given the correlations among the states, since the rows would not be linearly independent. By building this approximation of California pre-intervention (before Prop 99), we can approximate California post-intervention.

\begin{figure}[t]
\centering
\subfigure[synthetic control view]{\includegraphics[width=1.65in, angle=0]{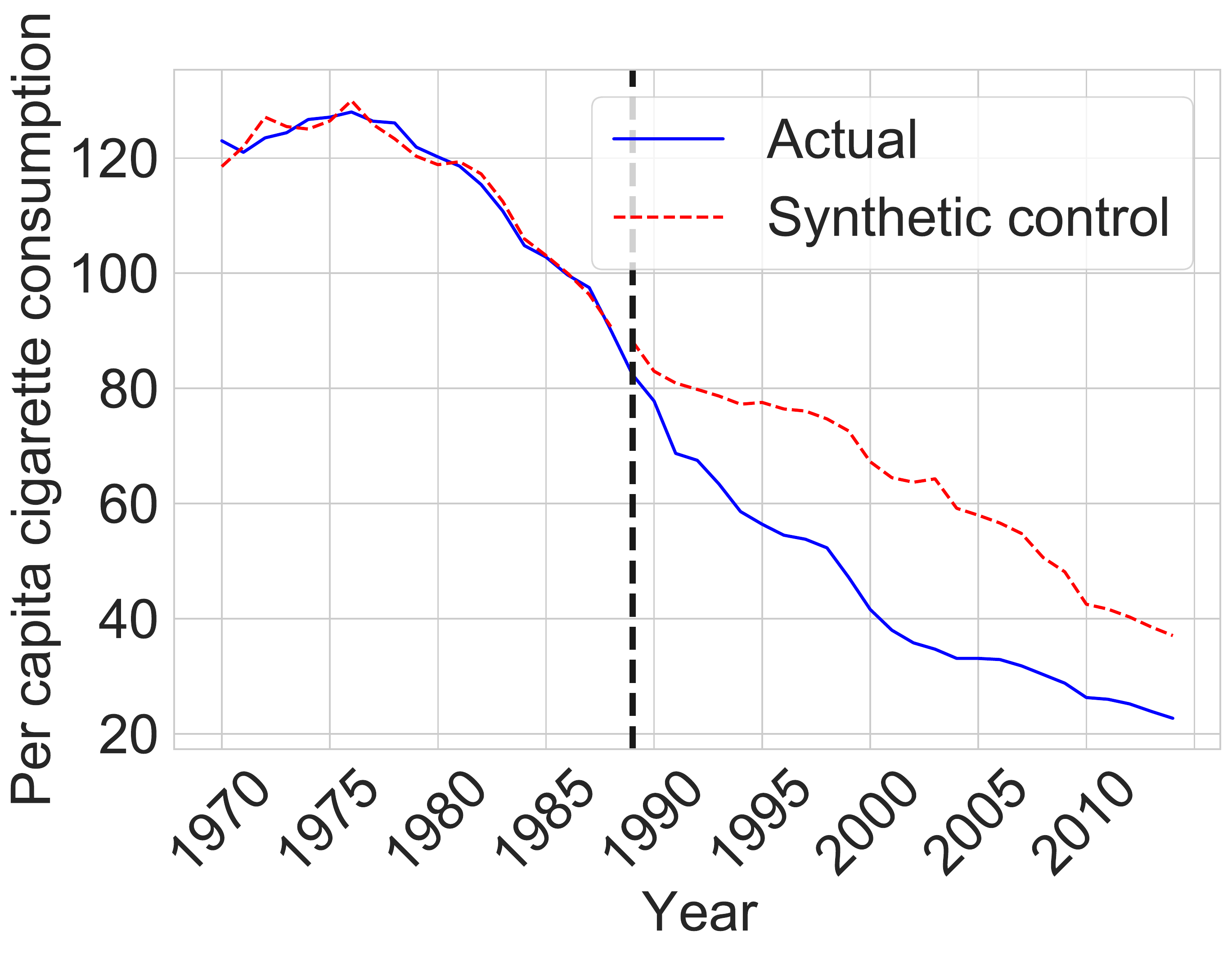}}
\subfigure[conventional A/B testing view]{\includegraphics[width=1.65in, angle=0]{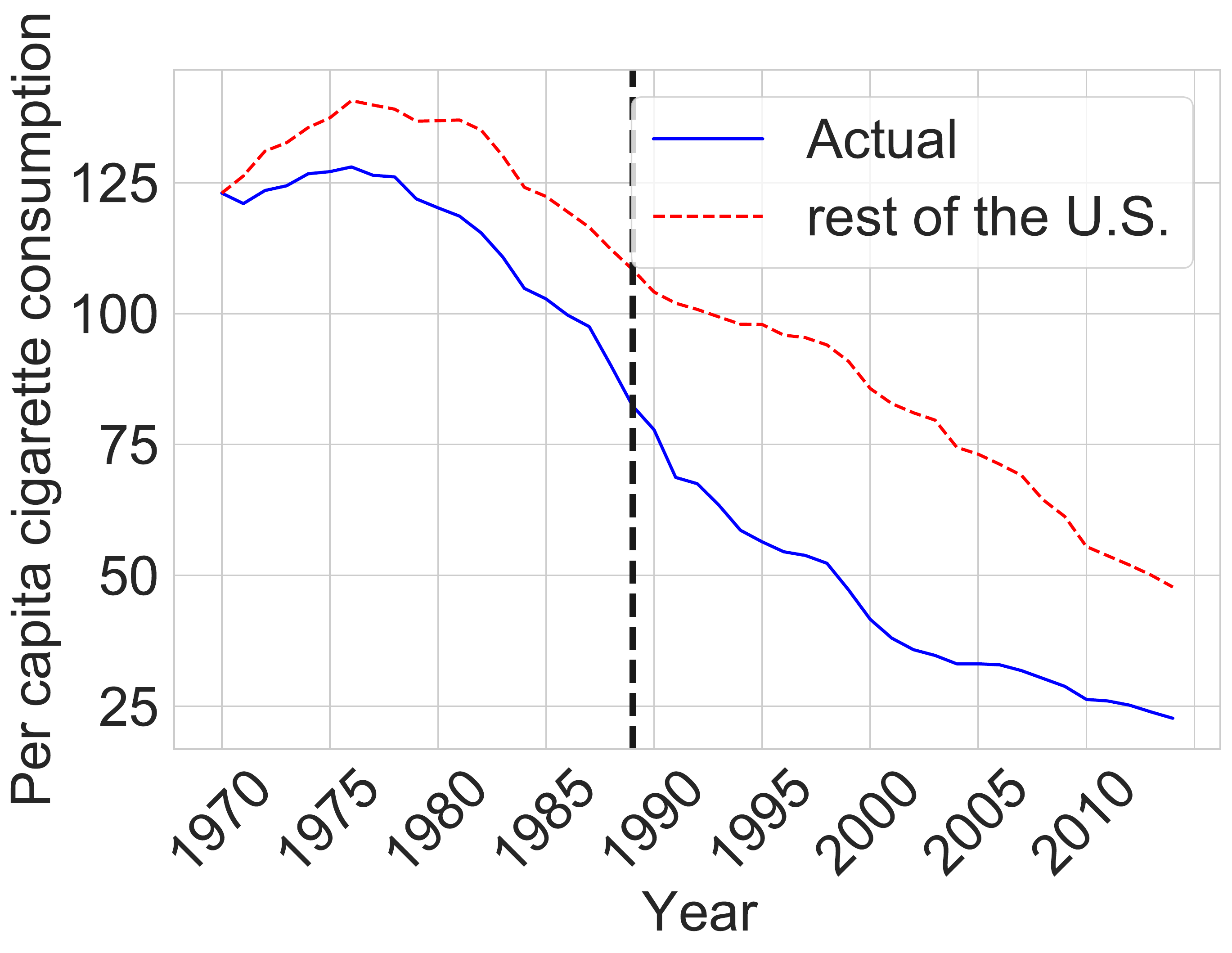}}
\caption{synthetic control versus A/B testing view for the state of California before and after Prop 99.}
\label{fig:prop99}
\end{figure}

Figure \ref{fig:prop99} shows both the synthetic control view and A/B testing view of the impact of Prop 99. The vertical dashed lines represent the time when Prop 99 was passed. The solid curve represents the raw data of California, and the dashed curves are the synthetic models building California from the rest of 49 states. In the synthetic control view, unlike the Conventional A/B testing view, the dashed line closely follows the solid line before Prop 99. Synthetic control gives a more precise and informative answer to the counterfactual ``What if there was no Prop 99''. Hence, in the synthetic control model we observe that the per-capita cigarette sales decrease more sharply in the actual California as opposed to the synthetic one. Therefore, we can claim that Prop 99 did have an impact on smoking.

For any controlled trial, we can ask and answer questions ``what is the impact of this change more precisely. In this work, we analyze the impact of lockdowns on the spread of COVID-19. We answer questions about the start of lockdown such as ``what if there was no lockdown'' or ``what if schools were closed earlier'', and in case of opening up the states, ``what if they were opened a few days or a few weeks earlier or later''. 



If the treatment unit is in its evolution, but the donor pool is ahead in its evolution, then we can predict how the treatment unit is going to behave. This way, we may build synthetic control models for both counterfactuals as well as predictions. In this work, we build a model to achieve both by analyzing COVID-19 data.

\subsection{Robust Synthetic Control}\label{sec:robust}

The robust synthetic control method is a generalization of the classical synthetic control method \cite{Amjad}. It makes the synthetic control estimation robust to randomly missing data and high variance noise. This generalization estimates the synthetic control weights using the unobserved mean values instead of the noisy observations. This estimation is done by ``de-noising’’ the data matrix using matrix completion and then using least squares regression to determine the synthetic control weights. It is a computationally efficient method with intuitive theoretical guarantees. Furthermore, the counterfactual outcome can be estimated by \emph{any} linear combination of the donor units, relaxing the convex constraints on the weights in classical synthetic control.

\paragraph{Data and model}
\label{sec:data}
Our observation matrix $\mathbf{X}$ looks like the following:
$$
\mathbf{X}=[X_{it}]=
\left[
\begin{array}{c|c}
D_{pre} & D_{post} \\
\hline
I_{pre} & I_{post}
\end{array}
\right],
$$
where $X_{it}$ denotes the metric (e.g., deaths) at location ${i}$ at time ${t}$. $D_{pre}$ is a matrix of the donor pool pre-intervention death counts, $D_{post}$ is a matrix of the donor pool post-intervention death counts, $I_{pre}$ is a matrix of intervention units pre-intervention donor-pool metrics and finally $I_{post}$ is the matrix of intervention units post-intervention metrics. 

As in \cite{Amjad}, we assume 
\begin{equation}
 \label{eq:resp_model}
 X_{it} = M_{it} + \epsilon_{it},
\end{equation}
where $M_{it}$ is the deterministic mean while the random variables $\epsilon_{it}$ represent zero-mean noise that are independent across ${i,t}$. Based on the theory of latent variable models \cite{usvt, LeeLiShahSong16, aldous, hoover1, hoover2}, we can state 
\begin{equation}
 \label{eq:latent_model}
M_{it} = f(\theta_{i},\rho_{t}) ,
\end{equation}
where $\theta_{i} \in \mathbb{R}^{d_1}$ and $\rho_{t} \in \mathbb{R}^{d_2}$ are latent feature vectors capturing unit and time specific information, respectively, for some $d_1, d_2 \ge 1$; the latent function $f : \mathbb{R}^{d_1}\times \mathbb{R}^{d_2} \rightarrow \mathbb{R}$ captures the model relationship. 

The treatment unit obeys the same model relationship during the pre-intervention period. That is, for $t < T{_0}$,
\begin{equation}
 \label{eq:model_intervention}
 X_{1t} = M_{1t} + \epsilon_{1t},
\end{equation}
where $M_{1t} = f(\theta_{1},\rho_{t}$) for some latent parameter $\theta_{1} \in \mathbb{R}^{d_1}, \rho_{t} \in \mathbb{R}^{d_2}$. If a unit had no intervention then equation (\ref{eq:latent_model}) would apply to both pre- and post-intervention periods. ~\cite{Amjad} indicates that the outcome random variables for all donor dispatch areas follow the model relationship defined by equations (\ref{eq:model_intervention}) and (\ref{eq:resp_model}). Therefore, the ``synthetic control” would ideally help estimate the underlying counterfactual means $M_{1t} = f(\theta_1,\rho_t)$ for $T_0 \le t \le T$ by using an appropriate combination of the post-intervention observations from the donor pool. In other words, $M_{1t}$ are the estimated metrics if there was no intervention applied to the treatment unit.

The last step in the process is to obtain the final synthetic control, i.e., the weights $\beta_i$'s. To that end, we assume that the mean vector of the treatment unit over the pre-intervention period, i.e., the vector 
 $M^-_1 = [M_{1t}]_{t \le T_0}$ , 
lies within the span of the mean vectors within the donor pool over the pre-intervention period, i.e., the
span of the donor mean vectors $M^-_i = [M_it]_{2\le i \le N ,t \le T_0}$. More precisely, we assume there exists a set of weights $\beta^\ast \in\mathbb{R}^{N-1}$ such that for all $t \le T_0$, $M_{1t} = \sum_{i=2}^{N} \beta_{i}^{\ast} M_{it}.$
The $\hat{\beta_i}$'s can be obtained as an estimate of true $\beta^*_i$'s via a process of regression on the de-noised version $\hat{\mathbf{M}}$ of the observation matrix $\mathbf{X}$, and they are the synthetic control weights.

\section{Methods}\label{sec:methods}

\subsection{Data preprocessing for COVID-19 spread}
In the case of pandemics, unlike synthetic control applications such as Prop 99, where everything happens at the same time in different regions, the virus becomes active at different times in different places. For instance, the COVID-19 spread started during November in Wuhan, but in New York, it hit in January, and in India, it arrived even later. Since the virus reaches different places at different times, the curves for the number of cases/deaths would also be at different stages of evolution at a given time.

\begin{figure}[t]
\centering
{\includegraphics[width=0.7\linewidth, angle=0]{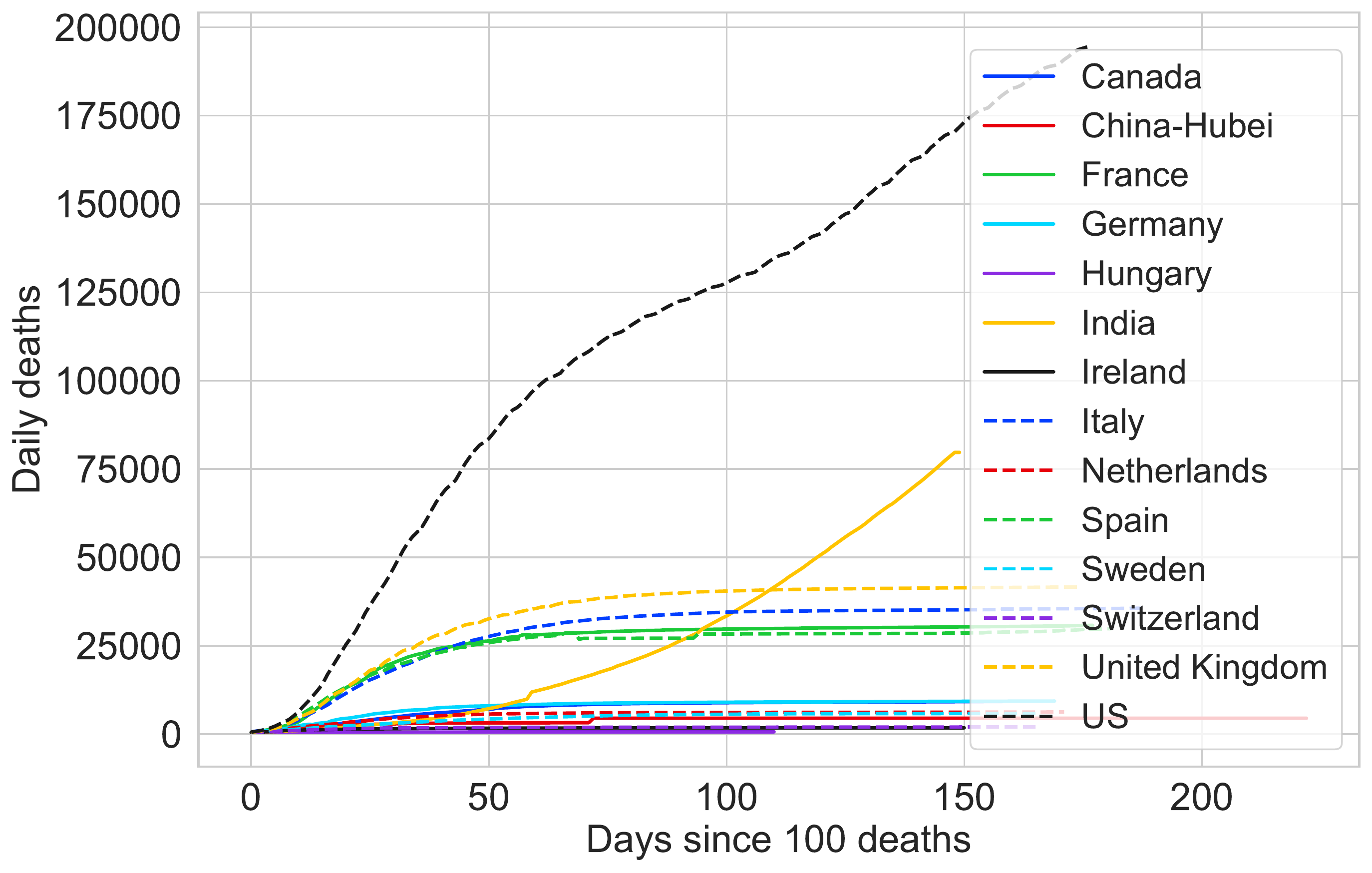}}
\caption{the number of deaths in different regions since $100$ COVID-related deaths occurred in each region. }
\label{fig:aligned_100deaths}
\end{figure}

To find correlations among different regions for the spread of COVID-19, we first align the timelines. One idea is to start tracking each region when it hits a threshold of deaths/cases (either absolute or population adjusted), and then assume the day the region hits that threshold is $t=0$. Different regions are now ``aligned'' on when the virus was active there. Figure \ref{fig:aligned_100deaths} depicts the number of deaths in different regions, where the x-axis represents the relative timeline in each region measured by the number of days passed since $100$ COVID-related deaths happened.

We use this information to build synthetic COVID-19 models. First, we pick a donor pool of regions where COVID-19 spread has been active for a number of days larger than some threshold, e.g. $30$ days. We then pick target areas where the spread has been active less than the original threshold, and more than a lower threshold, e.g., $15$ to $30$ days. After that, we build a model using the donor pool for a low threshold number of days ($15$ days) and do predictions for the rest of the interval ($15-30$ days) in the target areas. Since for some target areas we have data for more than $15$ days, we can compare the synthetic results to the actual data for those regions. In other words, we train the model using $15$ days of data and test the model using the rest $15$ to $30$ days.

However, the problem with this alignment with thresholds is that it does not account for control measures such as lockdowns, social distancing, etc. On the other hand, classical epidemiology models depend heavily on $R0$, the infection rate, which defines how quickly the disease spreads. All the control measures to ``flatten the curve'', such as social distancing, lockdowns, etc. reduce $R0$. The goal to design these control policies is to reduce $R0$ below $1$ so that the case becomes an ``endemic'' and the case numbers exponentially go down to $0$. In other words, if $R0 > 1$, the number of cases will continue to increase, changing the evolution of the cases/deaths curve.

A refined idea to address this issue is to pick $t_0$ not based only on thresholds, but based on the dates where lockdowns were put in place in each region. We used data from IHME (Institute for Health Metrics and Evaluation) with different publicly identified lockdown measures for different places \cite{IHME}. These measures include the dates where restrictions where imposed, such as mass gathering restrictions, stay at home order, educational facilities closure, initial business closure, nonessential services closure, etc. We pick the strictest lockdown date as $t_0$ since for the accuracy of our model we need $R0$ to be constant (and minimum) from then on. To our knowledge, the strictest lockdown date would be the date of the last announced measures.

\begin{figure*}[t]
\centering
{\includegraphics[width=\textwidth, angle=0]{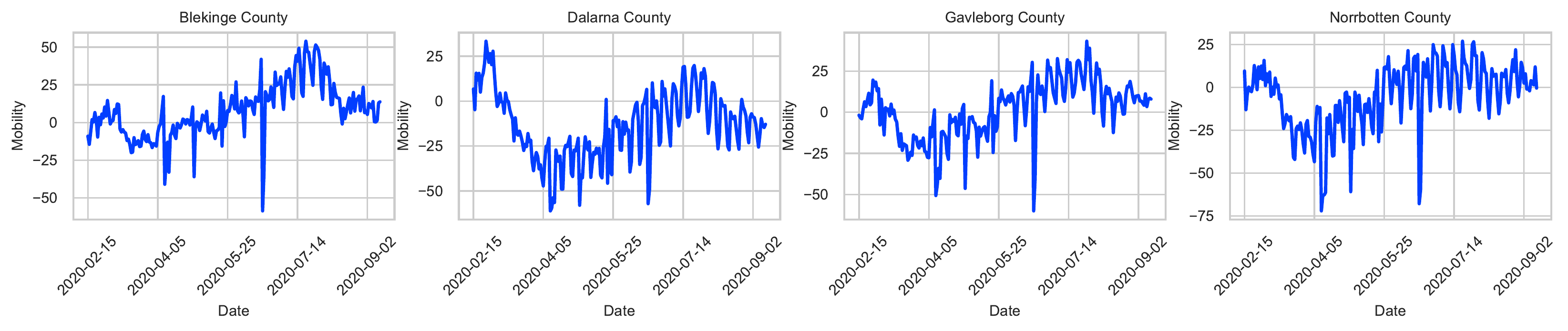}}
\caption{The average mobility reported by Google for four different Swedish counties during spread of COVID-19. }
\label{fig:google_mobility}
\end{figure*}

To complete this dataset and account for regions that are not measured in this data, we augmented it with the mobility data provided by Google and Apple regarding how people's mobility has changed during the lockdown. This data is gathered from people's smartphones and assumes normal conditions as the base-level, where the average mobility of people is around $0$. When this data drops from $0$, it implies that lockdowns may have been imposed in a given region and as a result, the average mobility has dropped. Figure \ref{fig:google_mobility} depicts the mobility changes in four counties from Sweden - even though Sweden did not formally have a lockdown, the spread of COVID caused people to voluntarily reduce their movement, leading to an implicit lockdown.

\begin{figure}[t]
\centering
\subfigure[relative date since lockdown]{\raisebox{6mm}{\includegraphics[width=1.65in, angle=0]{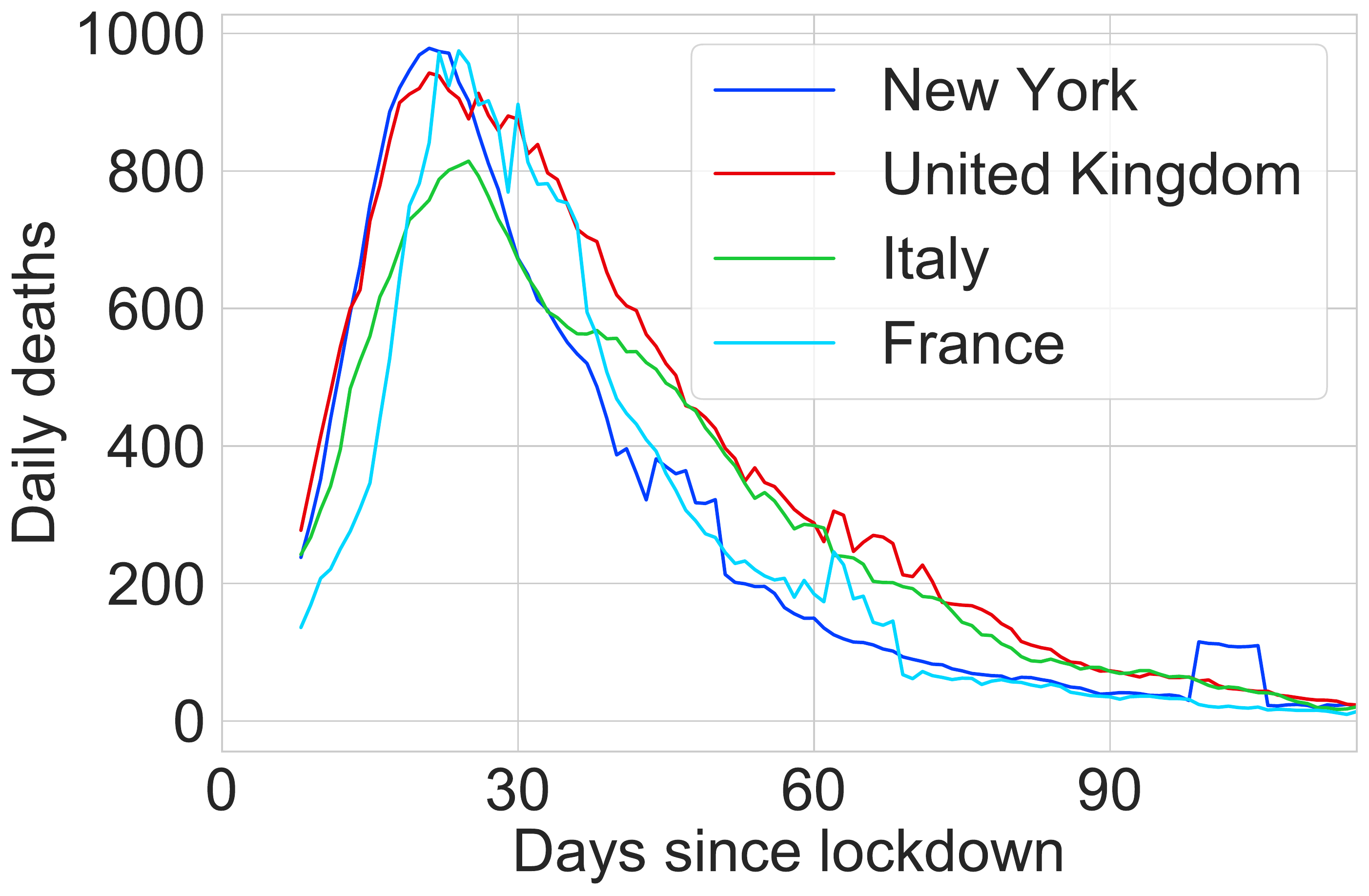}}}
\subfigure[chronological date]{\includegraphics[width=1.65in, angle=0]{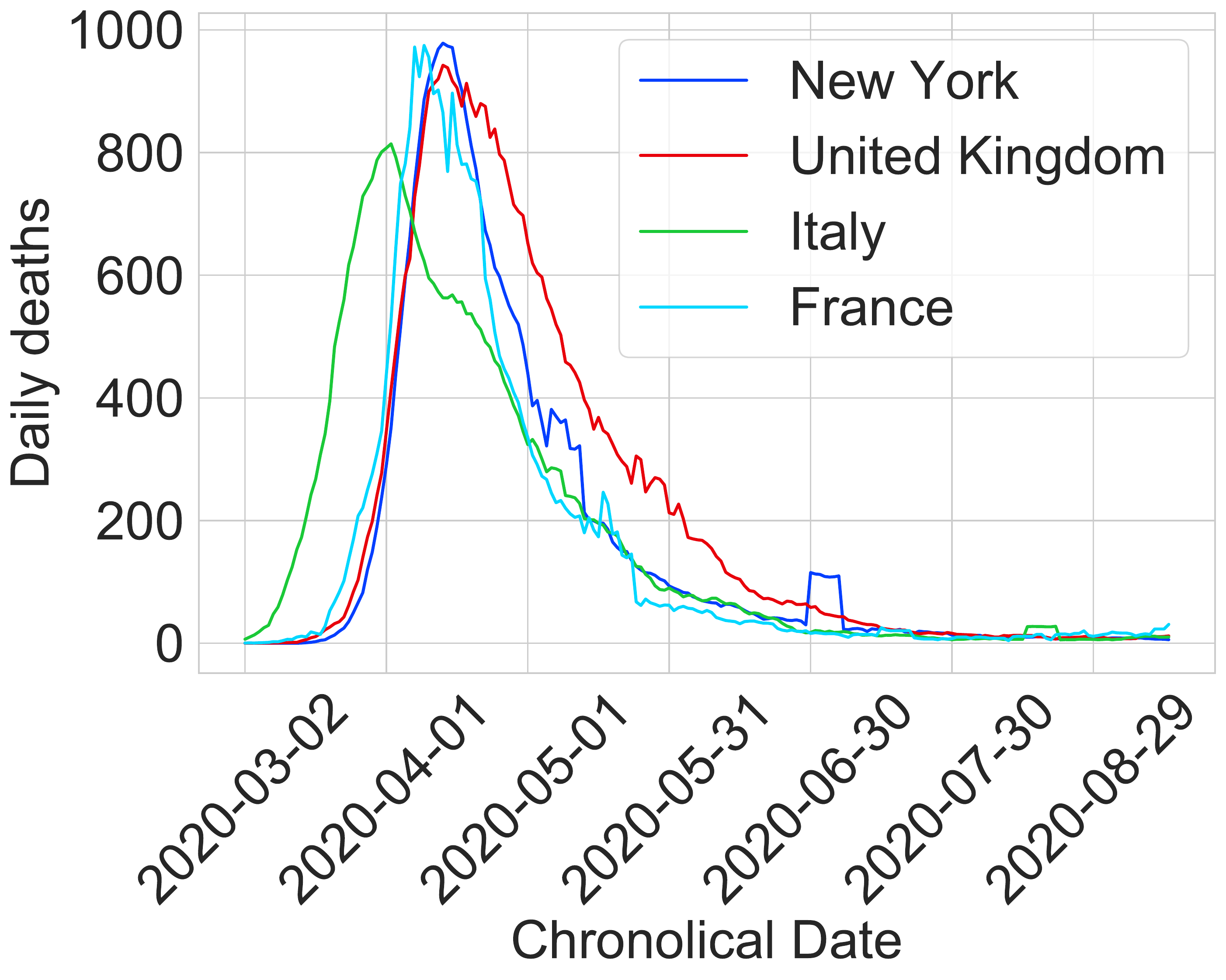}}
\caption{The moving average of the number of deaths in a) relative days since lockdown b) chronological order}
\label{fig:align_vs_chron}
\end{figure}

Figure \ref{fig:align_vs_chron} depicts the moving average for the number of deaths in four regions. The graph on the left is time-aligned based on the start of lockdown in each region, and the one on the right depicts the chronological data. We observe that in the aligned data, a clear pattern emerges and the number of death in those regions are very similar. The reason why these regions are selected to be presented is that the synthetic control model found a high correlation in the number of deaths among them. 

\subsection{Impact of lockdowns}

\begin{figure}[H]
\centering
{\includegraphics[width=0.7\linewidth, angle=0]{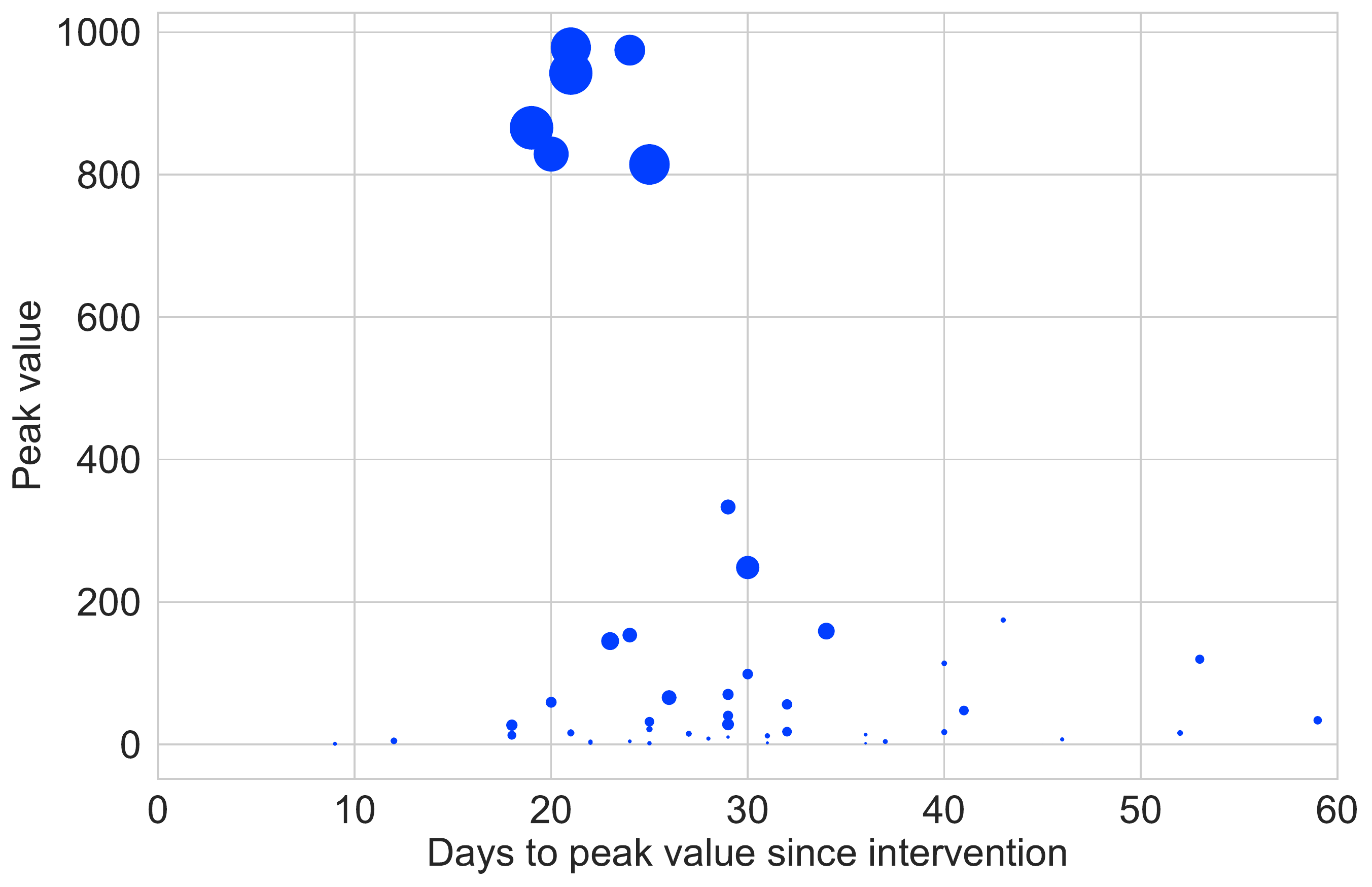}}
\caption{the impact of lockdown in different regions, measured by death count when interventions are applied.}
\label{fig:lockdown_impact}
\end{figure}

The strictness of lockdown and the time it is applied in different regions has a major impact on the spread of the virus. Figure \ref{fig:lockdown_impact} depicts the impact of lockdown using a scatter plot, where each point represents a region. The $y-axis$ represents the peak value for each region (maximum number of daily deaths in this case), and the $x-axis$ is the number of days since the intervention applied until the region has reached its peak. Also, the size of each marker is proportional to the number of deaths in each region on the day lockdown was imposed. 

We observe that first, the higher number of deaths are on the day of the lockdown, typically, the peak value is higher, and second, for places with high death counts, the peak is reached in roughly three weeks. This phenomenon was consistent, especially in the early days of the spread of COVID-19, with the typical evolution of a fatal case - from infection to incubation to death took roughly three weeks. In places with high fatalities (the markers on top are Western European countries and New York), this indicates that the lockdowns were strict and effective in immediately reducing the spread of infection down. In other places where the fatalities were not that high on the date of the lockdown, the infection continued to spread for several days, and the peak happened several days after three weeks, indicating a somewhat lax lockdown.

\section{Results}\label{sec:results}
In this section, we start by presenting how our synthetic control model works to predict the spread of the disease in different regions, as well as counterfactual and synthetic intervention analysis. The counterfactual analysis aims to show how the evolution of the disease would have looked had the lockdown measures put in place earlier or later. Furthermore, using counterfactuals and synthetic interventions, we show how different policy measures impact the spread of the disease.

\subsection{Predictions}
We present an example of applying synthetic control to predict how the spread of COVID would look in New York and compare it to how it actually looked in New York. Furthermore, we do a counterfactual analysis of the impact of lockdown imposing dates. New York (City) had the closest correlation both in terms of demographics as well in the spread of the virus with Western European countries in the early days, and we use Western Europe as the donor pool. Additionally, besides the similarity in the way the pandemic was initially handled and the lockdowns put in place, it was observed that the strain of the virus that spread in New York came from Western Europe (primarily Italy), rather than China~\cite{Gonzalez-Reiche297}.

\begin{figure}[t]
\centering
\subfigure[synthetic control prediction for NY cases]{\includegraphics[width=\linewidth, angle=0]{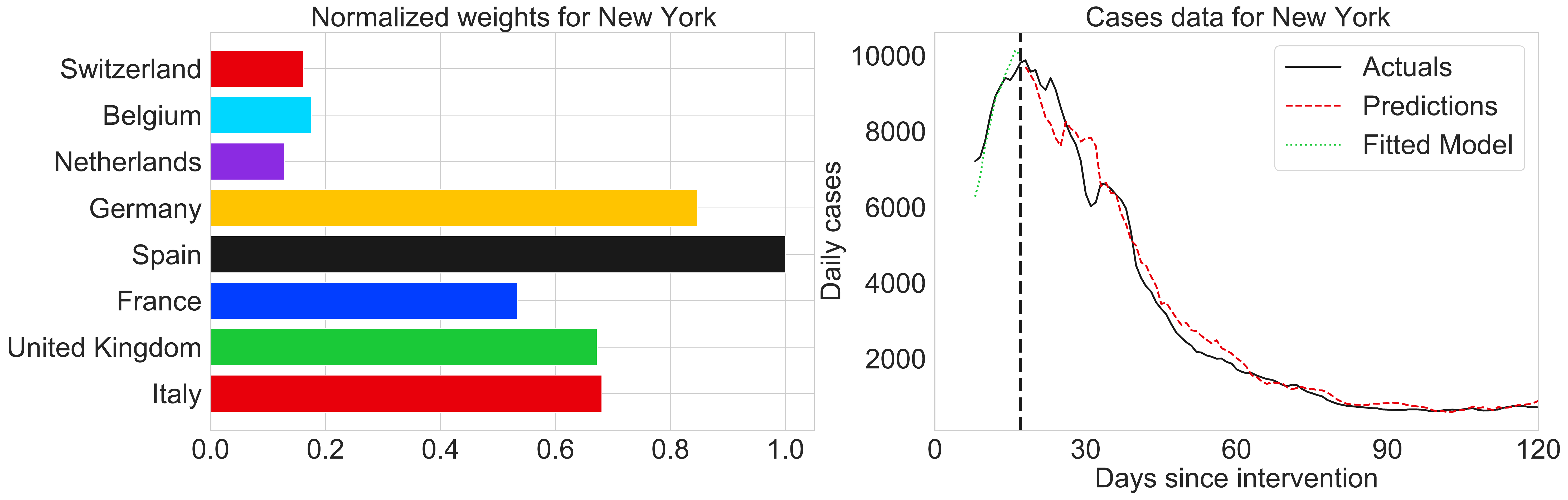}}
\subfigure[synthetic control prediction for NY deaths]{\includegraphics[width=\linewidth, angle=0]{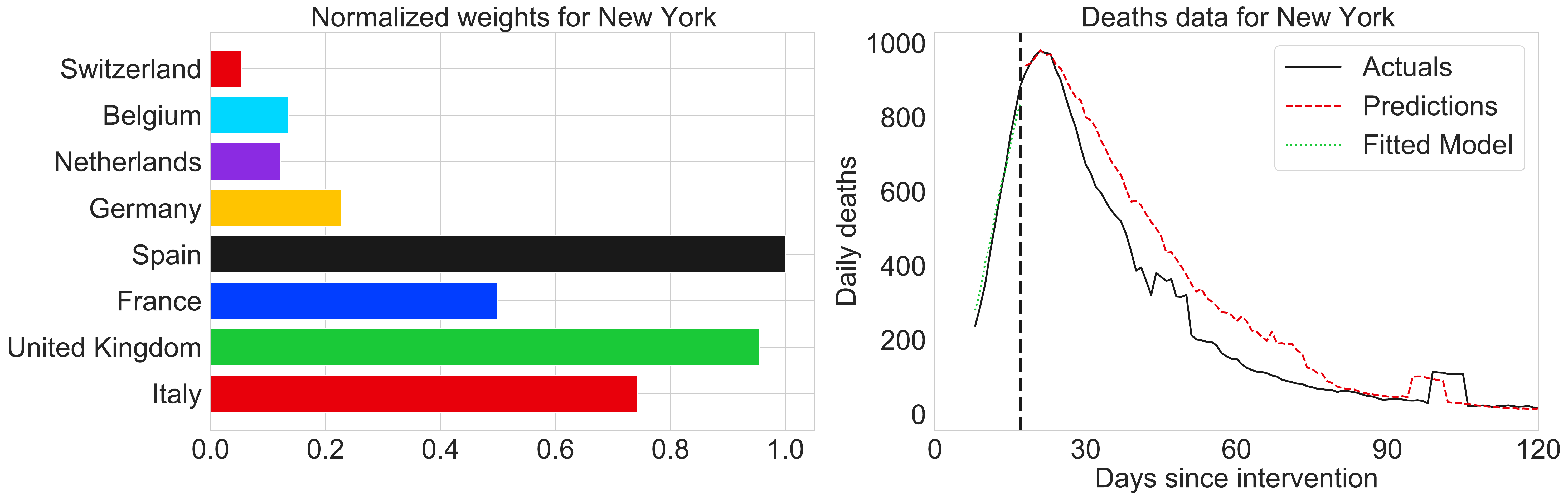}}
\caption{Prediction of NY cases and deaths versus reality. The model has been trained for $20$ days and used for $\approx 100$ days of prediction. }
\label{fig:cases_deaths_NY}
\end{figure}

Figure \ref{fig:cases_deaths_NY} depicts the number of cases and deaths in New York using our synthetic control model. The donor pool is selected from Western Europe, and the graphs on the left show the normalized weight of each region, where $8$ regions with the highest weights are selected to be shown.

We observe that the countries where COVID-19 hit more heavily have higher weights in the synthetic control model as opposed to the countries that managed to keep the number of cases low. Furthermore, note that a bump which is observed in the number of deaths after day $90$ comes from the data source and is due to the different measurement methods applied on those days, which since occurs in all the regions, the synthetic control model predicts it as well, and the shift is because Western Europe was mostly ahead of New York during the spread.

\subsection{Impact of lockdown: counterfactual analysis}
\begin{figure}[t]
\centering
{\includegraphics[width=\linewidth, angle=0]{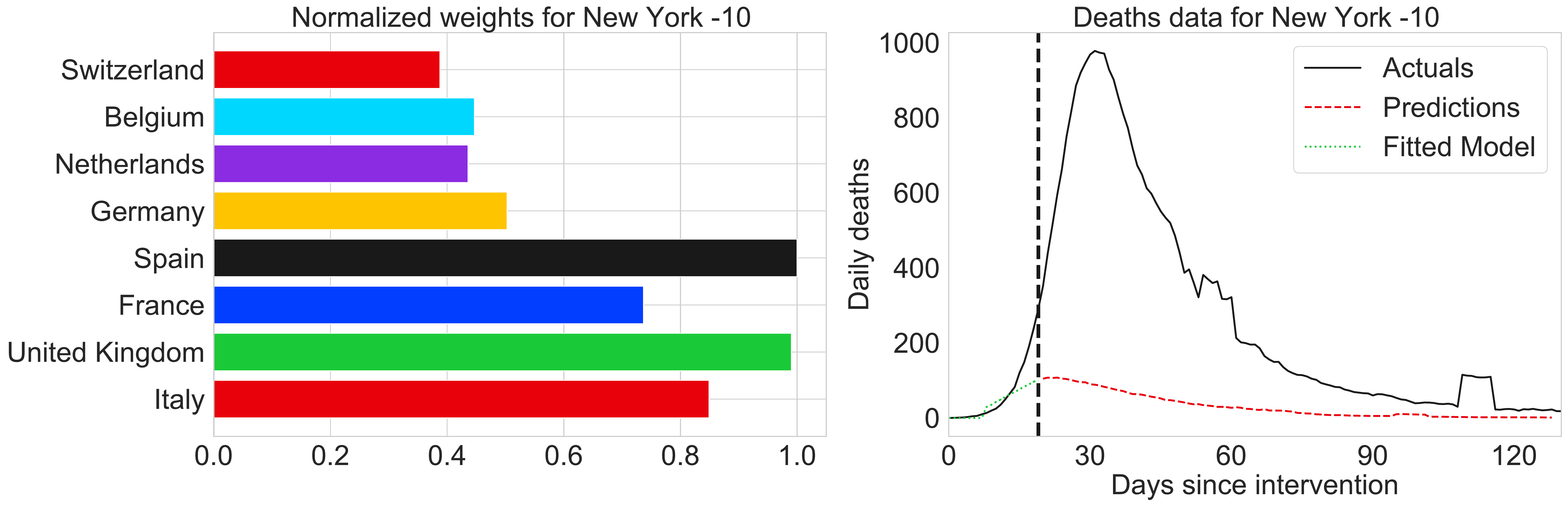}}
\caption{NY deaths counterfactual prediction if the lockdown was done $10$ days earlier, versus actual NY deaths. }
\label{fig:deaths_NY-10}
\end{figure}

Figure \ref{fig:deaths_NY-10} depicts how NY cases would have looked if the lockdown had been imposed $10$ days earlier, with the same donor pool as Figure \ref{fig:cases_deaths_NY}. We observe that if lockdown was imposed earlier in NY, the number of cases would have decreased considerably. Note that compared to Figure \ref{fig:cases_deaths_NY}(b), the weight of countries in which COVID-19 spread was more controlled has increased since they have started the lockdown early on during the spread, and if NY had locked down only $10$ days earlier, its spread pattern would have been more similar to those countries, with a \emph{significantly} reduced death count. While the precise values are not important and our results should be taken more qualitatively, our models show that deaths in New York could have been reduced by over 80\% by pushing the lockdown 10 days ahead of when it was finally done.

\begin{figure}[t]
\centering
\subfigure[synthetic control prediction for NY cases]{\includegraphics[width=\linewidth, angle=0]{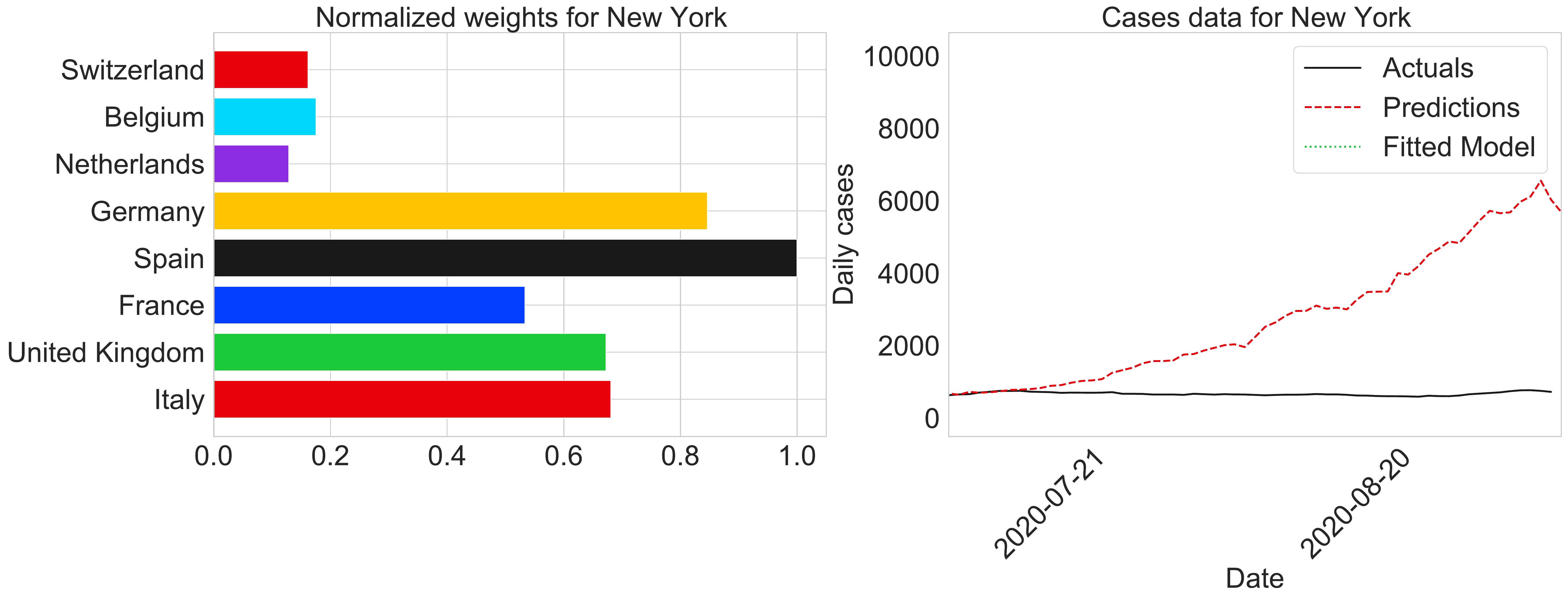}}
\subfigure[synthetic control prediction for NY deaths]{\includegraphics[width=\linewidth, angle=0]{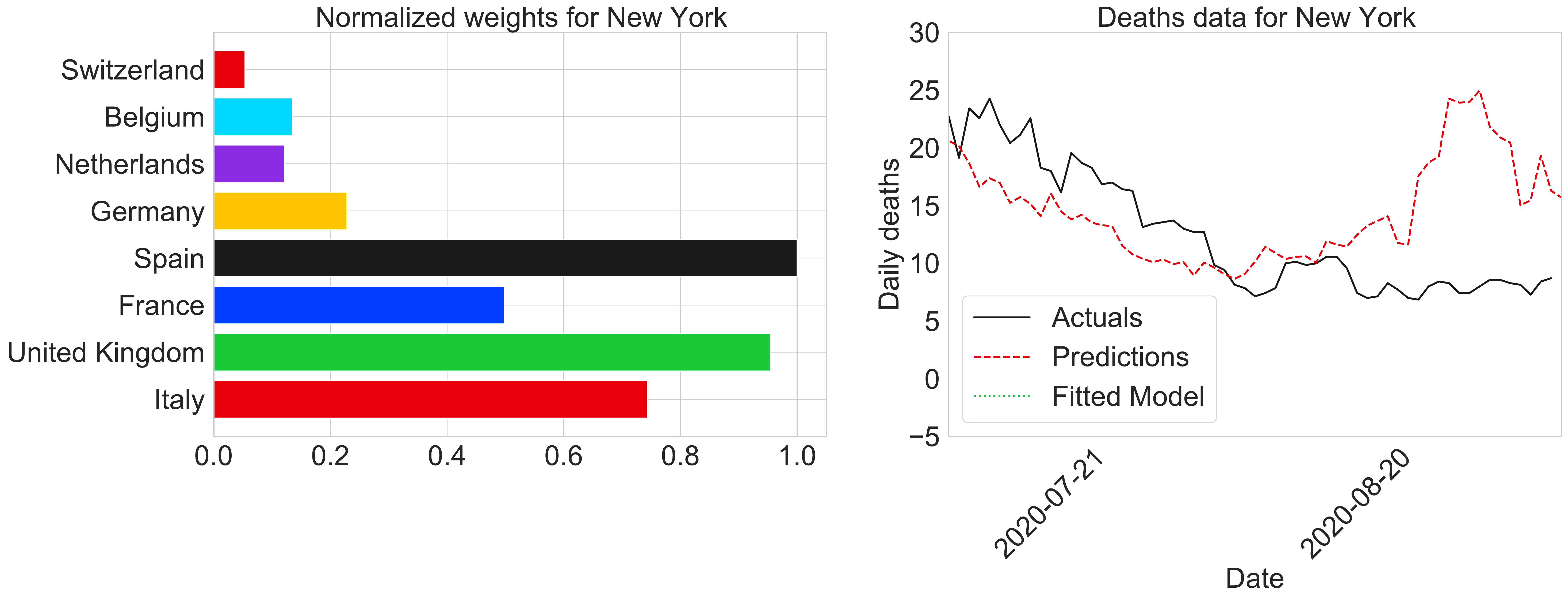}}
\caption{Post Memorial-Day Counterfactual predictions of NY cases and deaths based on a Western European donor pool. NY is not experiencing a second wave, unlike Western Europe}
\label{fig:cases_deaths_NY_last60}
\end{figure}

Note that in performing predictions and building counterfactuals, it is important to pick the right set of donor pool elements. While in the early phase conditions matched between New York and Western Europe, things have looked different in the summer of the northern hemisphere. Western Europe is undergoing a significant second wave, whereas it is largely under control in New York, and this can be seen in Figure~\ref{fig:cases_deaths_NY_last60} where the actual cases and deaths in New York are significantly lower than the counterfactual built using the Western European donor pool. One of the speculated reasons is that Western Europe opened up indoor dining and bars whereas they are still banned in New York. Also, consistent with the dynamics of the disease, deaths are lagging cases by roughly three weeks in Western Europe.
\subsection{Post-memorial day second wave in the US: likely causes}

In this section, we analyze how different regions differ in policies of temperature,  bars and dining and reopening and if there are any correlations between these factors and post-memorial day spread.

\subsubsection{Impact of post-Memorial Day temperatures and AC usage}
\begin{figure}[t]
\centering
\raisebox{8mm}{\includegraphics[width=0.5\linewidth, angle=0]{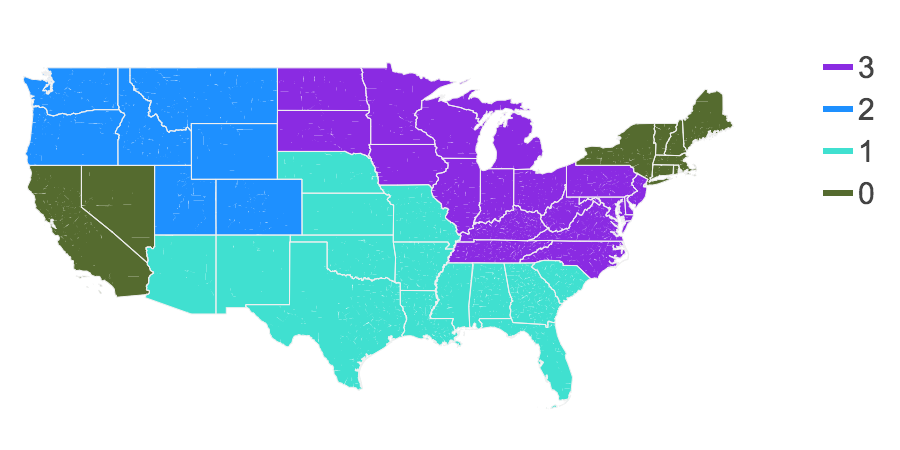}}
{\includegraphics[width=0.45\linewidth, angle=0]{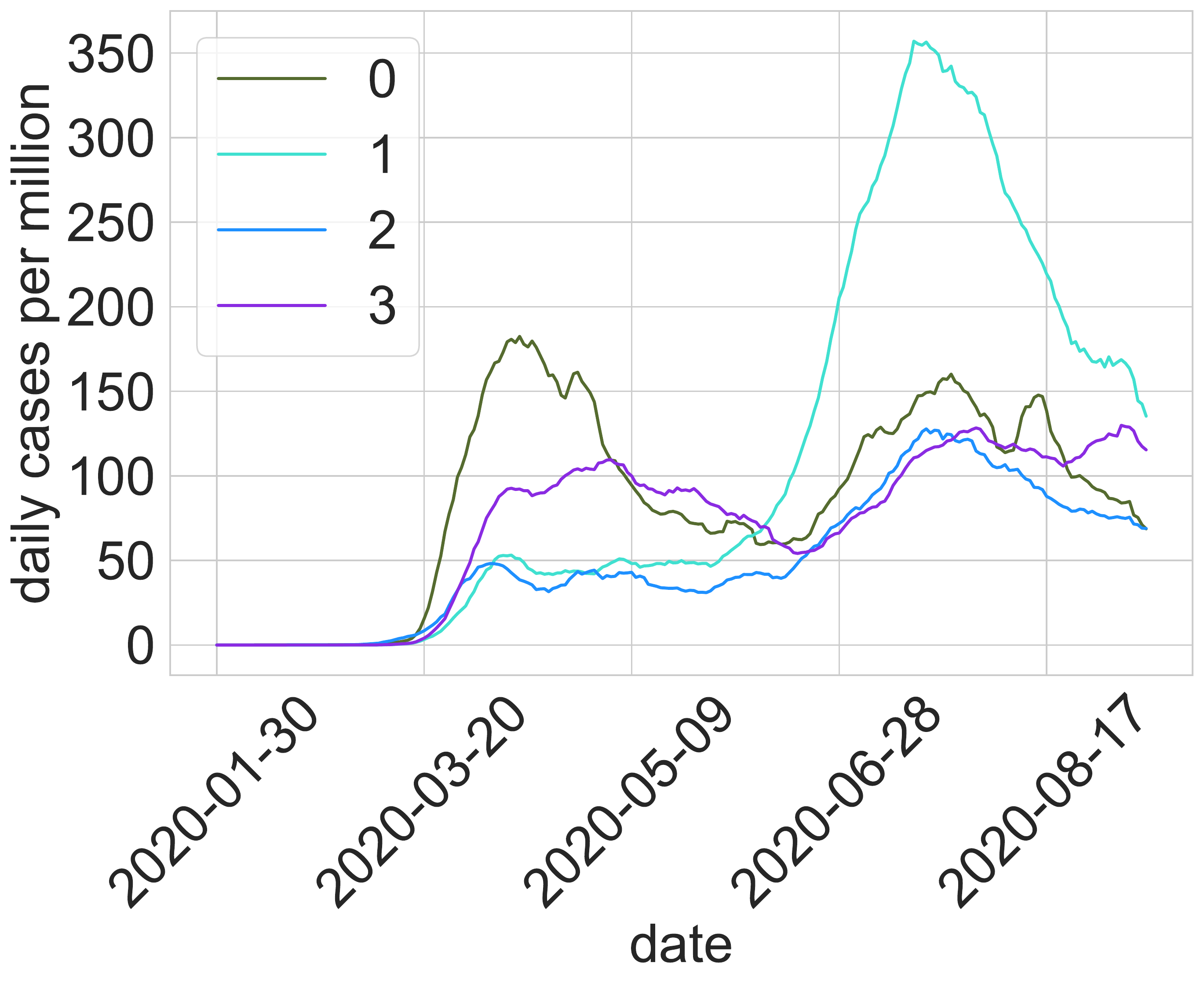}}
\caption{regions clustered according to post-memorial day temperature trends, and daily number of cases per million in each region.}
\label{fig:heat_cases}
\end{figure}

 One of the postulated reasons for the second wave in the US has been the summer heat and increased use of ACs, which cause the virus droplets to recirculate. We employ (simple k-means) clustering to divide the U.S. states into different regions, based on their daily temperature after memorial day. Figure \ref{fig:heat_cases} depicts that when the U.S. is divided into four regions based on temperature. The graphs show the daily number of cases per million, where the hottest states suffer the worst outbreak after memorial day. Especially in the southern regions (cyan) which is the hottest, we observe that number of cases per million continues to rise. However, all 4 regions displayed a second wave.

Table \ref{Tab:avg_tmp} shows the average temperature of each cluster, as well as the average number of daily cases and deaths per million for each, post-Memorial Day. The data in the table however indicates that there is no clear correlation in the temperature and case levels in the regions.

\begin{table*}[t]
\centering
\footnotesize
\begin{tabular}[t]{c c c c} 
\toprule
region & average temperature (F) & average daily cases per million & average daily deaths per million \\
\midrule
Cluster $0$ & $60.33$& $7984$& $136$ \\
\midrule 
Cluster $1$ & $75.74$& $16819$& $375$ \\
\midrule
Cluster $2$ &$65.96$& $12427$& $375$ \\
\midrule
Cluster $3$ & $69.78$& $11608$& $414$ \\
\bottomrule 
\end{tabular}
\caption{average number of daily cases and deaths per million in each region, post-Memorial Day}
\label{Tab:avg_tmp}
 \end{table*}



\subsubsection{Impact of Bars and Indoor Dine-In}

\begin{figure}[t]
\centering
{\includegraphics[width=0.7\linewidth, angle=0]{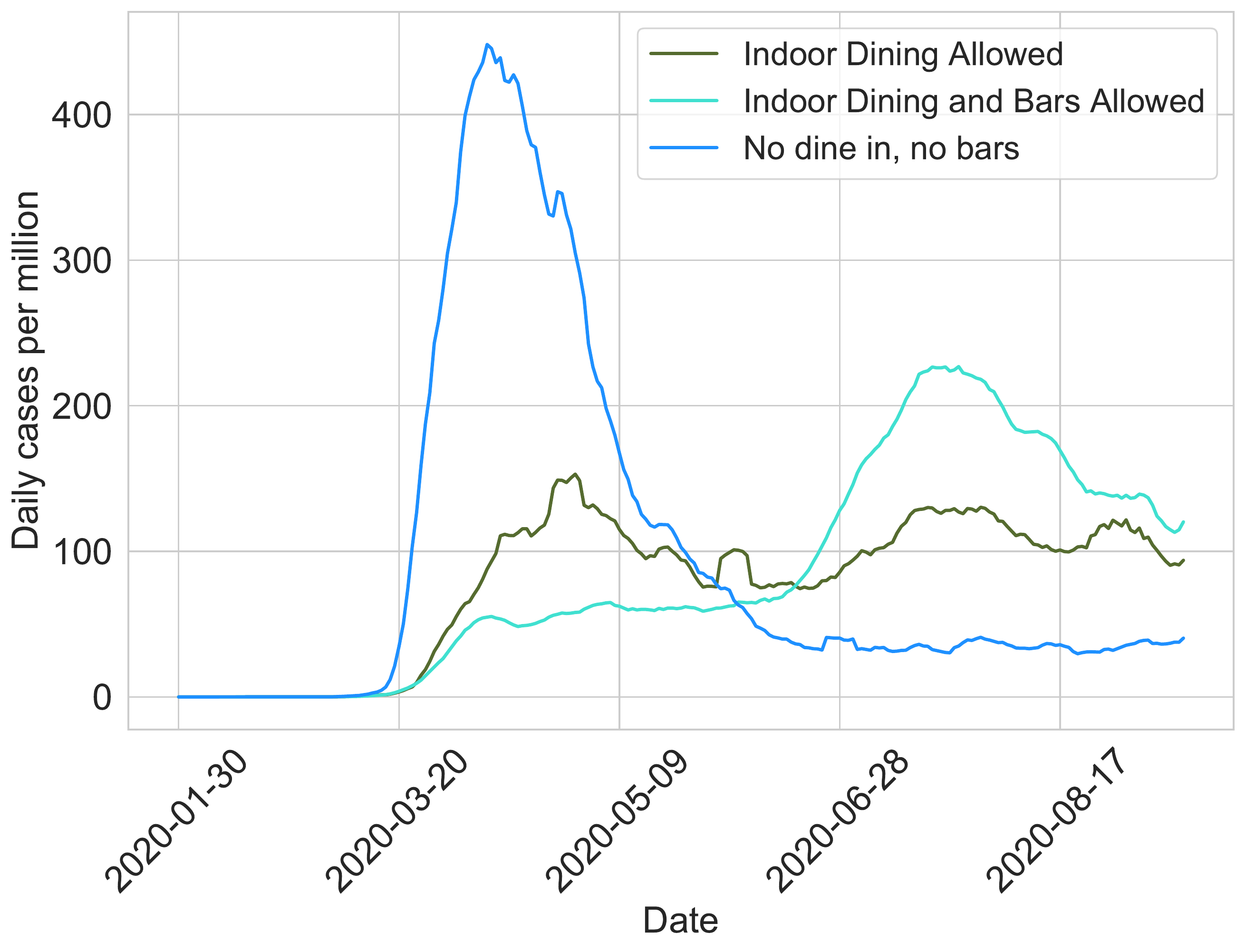}}
\caption{Number of daily cases per million in states with different indoor dining/bars policies. New York, Massachusetts, Connecticut, and New Jersey restrict most indoor activities. In New Mexico, Kansas, District of Columbia, Rhode Island, and North Carolina, indoor bar service was not allowed.}
\label{fig:indoor_bars}
\end{figure}

We use the data from \cite{eaterjun2020} to depict the impact of Bars and indoor dinings on the spread of the virus. New York, New Jersey, Massachusetts\footnote{Massachusetts allowed limited dine-in from late June, but most restaurants did not open up immediately \url{https://boston.eater.com/2020/6/22/21298937/indoor-dining-massachusetts-restaurants-june-22-guidelines} }, and Connecticut banned all indoor dine-ins, while in New Mexico, Kansas, District of Columbia, Rhode Island, and North Carolina, indoor bar service was not allowed. We group the rest of the states together as ones that allow indoor dining and bars. Figure \ref{fig:indoor_bars} depicts this impact on the three groups. 

We observe that in states where indoor dining and bars were not allowed, the number of cases reached to its maximum earlier than the rest of the states, and then decreased very quickly until it became much smaller than the number of cases in those. The early peak was due to the early arrival of the virus on those regions, which spread quickly before the lockdown policies were imposed. In the states where indoor dining and bars were allowed, the number of daily cases per million kept increasing until it became higher than the other two groups, even though the initial number of daily cases was lower. Therefore, we conclude that indoor dinings increase the spread of the virus, and indoor bars increase it even further\footnote{Note that the number of cases per million is a proxy of the density of cases, i.e., it is computed as per capita infection rate $\times 1,000,000$ }.



\subsubsection{Mobility Analysis}
We have clustered different regions according to how their mobility has changed after the lockdown. We used the mobility data corresponding to ``retail and recreation'' from Google mobility trends, and using the k-means clustering method, we revealed $4$ clusters representing how much and how soon the regions have opened up.

\begin{figure}[t]
\centering
{\includegraphics[width=\linewidth, angle=0]{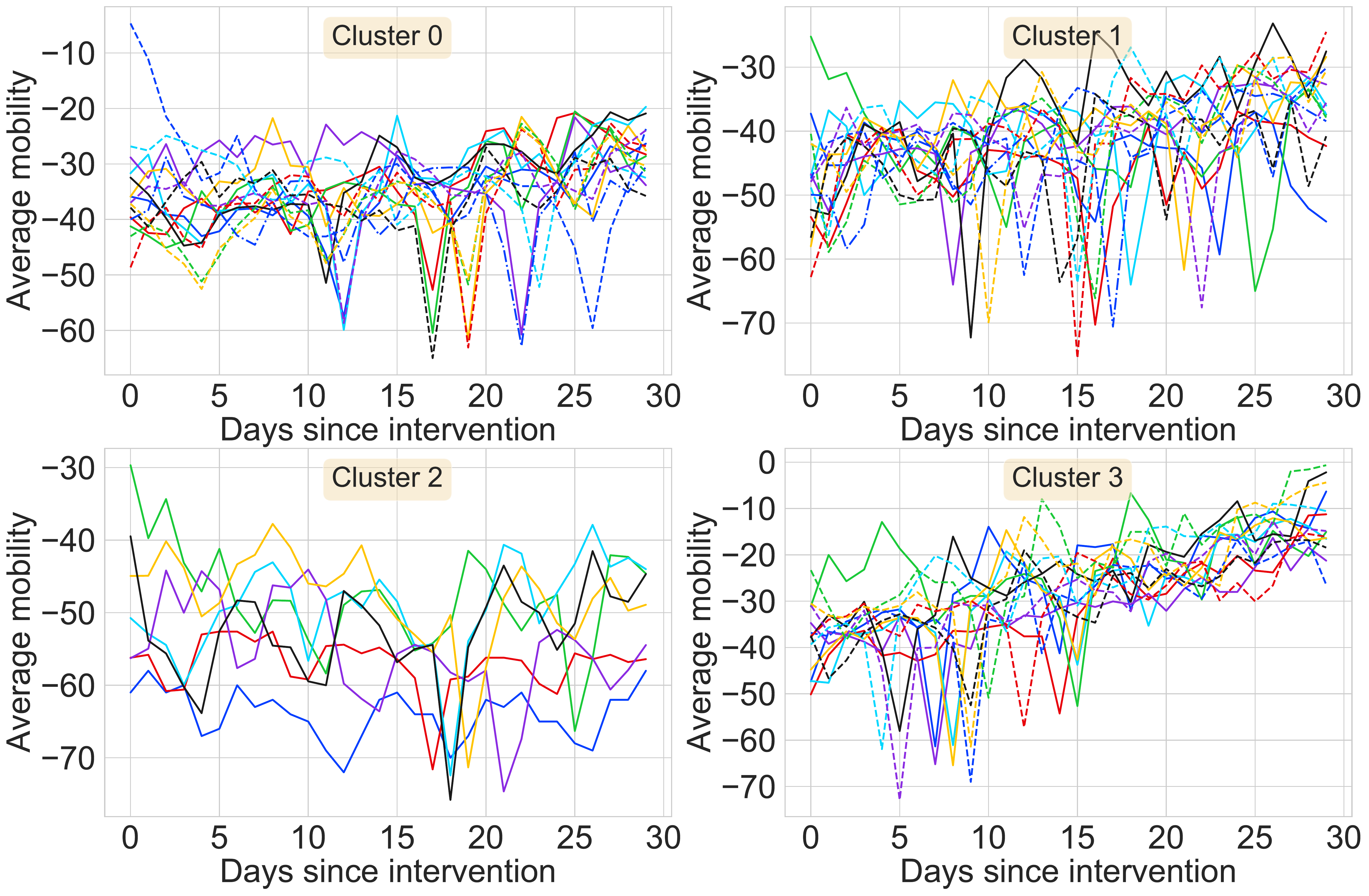}}
\caption{Four different clusters based on mobility trends of the US states after the lockdown. The trends are derived from Google's mobility data on `retail and recreation''.}
\label{fig:mobility_clusters}
\end{figure}

Figure \ref{fig:mobility_clusters} depicts the four different clusters with different mobility trends. In cluster $1$, the mobility goes back to normal fairly quickly after the lockdown, whereas in cluster $0$, the mobility stays low for a longer period of time. Furthermore, comparing clusters $1$ and $3$, we observe that the mobility keeps increasing and decreasing after the lockdown, respectively. Needless to say, there might be some regions clustered into a certain group by the clustering model, and may not be sensible by looking at the graphs. 

\begin{figure}[t]
\centering
\raisebox{8mm}{\includegraphics[width=0.5\linewidth, angle=0]{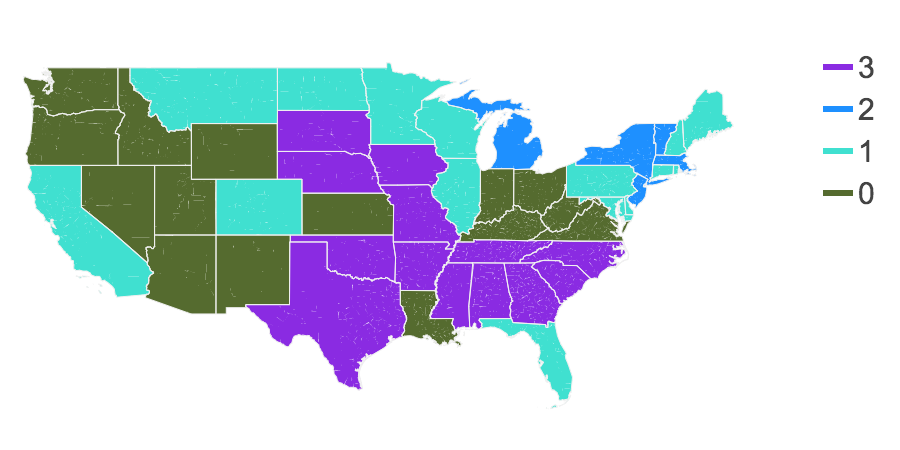}}
{\includegraphics[width=0.45\linewidth, angle=0]{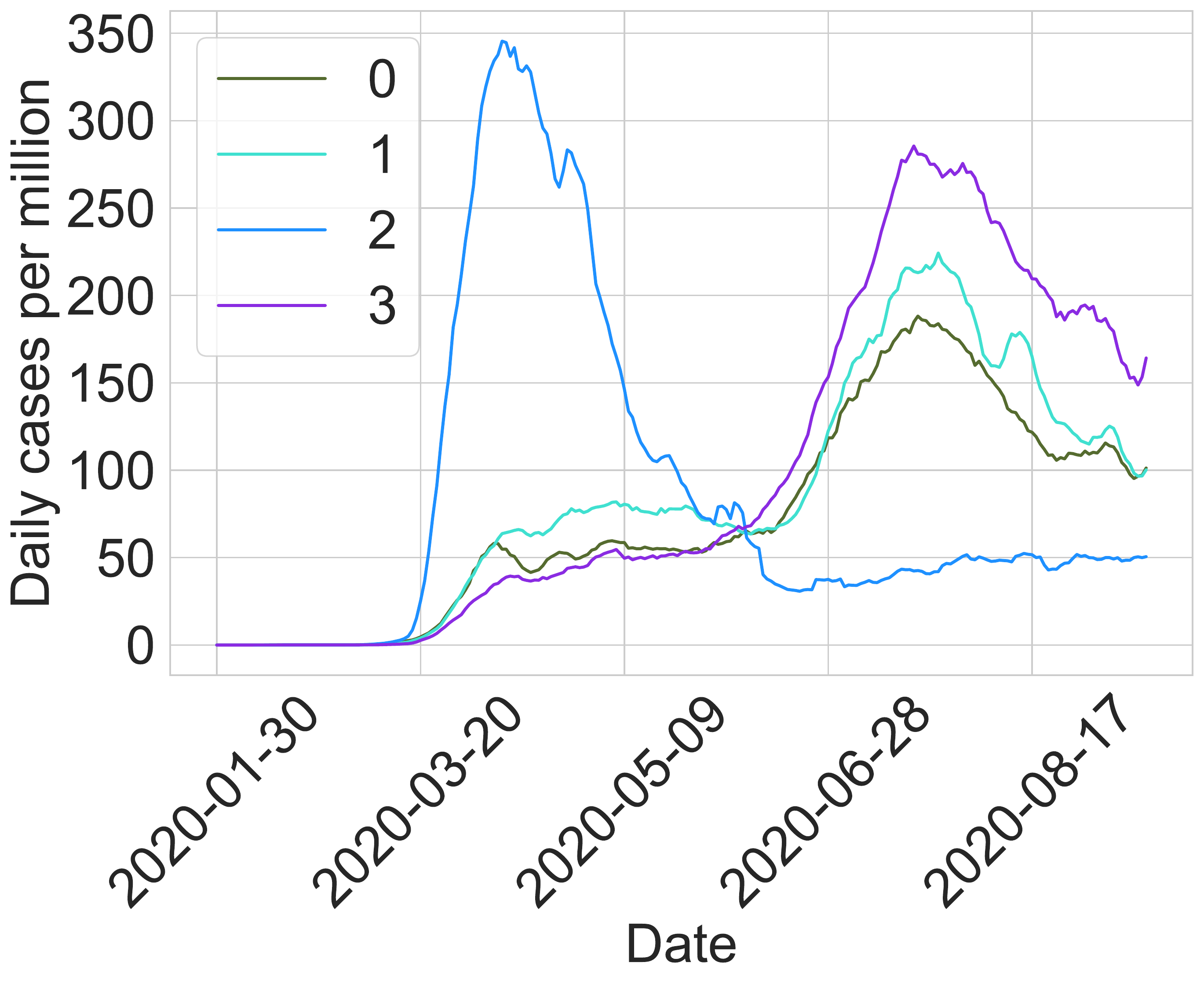}}
\caption{regions clustered according to post-lockdown mobility trends, and daily number of cases per million in each region.}
\label{fig:mobility_cluster_regions}
\end{figure}

Figure \ref{fig:mobility_cluster_regions} depicts the regions belonging to each cluster, as well as the moving average of their daily cases. We observe that in clusters $0$, $1$, and $3$, the number of daily cases continues to rise, and their peak happens much later than that of cluster $2$. However, in cluster $2$, the number of daily cases reaches to a maximum earlier and then begins to drop. In this figure, region $2$ represents the North-East minus Pennsylvania. However, comparing this to Figure \ref{fig:global_us_cases}, we observe that if we consider Pennsylvania in the North-East, the number of daily cases would be slightly higher, indicating the Pennsylvania shows a different behavior than the other states in the North-East with respect to mobility and opening up. 

\subsection{Herd Immunity vs. Active Interventions}
In different states of the United States, major outbreaks happened after memorial day. To analyze what has caused this behavior, we use the technique of Synthetic Interventions \cite{SyntheticInterventions} to analyze the data. We first set the intervention date to memorial day, then, we train a model for each county based on a donor pool of counties from a specific region. We picked ``North-East'' to be our donor pool which is New England +(New York and New Jersey)\footnote{Note that in the previous section almost every way we cut the data, the states that are in the North-East would belong to the cluster that had cases under control after Memorial Day.}. Then, we filter the donor pool to only include counties at ``similar stages'' of coronavirus spread, i.e. counties that had similar case numbers, when the population was adjusted. To do so, we first consider the number of cases per million in each region, and then we pick donor counties in similar range, within $50\%$ of the cases of the target county. Furthermore, to avoid overfitting, we use a low rank approximation where we use three singular values for the matrix that has a nominal rank at least $10x$ that number. Then, we compare the synthetic model of each county with its actual behavior.

One factor which might have played a role in our analysis is that since the spread started early in the North-East, enough people may be infected and have ``herd immunity'', which in turn has slowed down the virus spread. Classical epidemiological models like SEIR show a natural slowdown in the disease spread as the fraction of people susceptible goes down, and ``herd immunity'' as a concept is often discussed \cite{herdimmunity}. While the number is still being debated, scientists believe that when a certain fraction of the population ($20\%-70\%$ are the figures currently under debate) have already been infected and immune to the disease, the region will reach herd immunity. A natural question to explore given our dataset is that the different behaviors we observe can be simply attributed to "herd immunity" or do other factors like active interventions play a part in the spread of the disease. Our prior analysis has shown that the North-Eastern states have generally had stricter lockdown measures, whether it is the speed of reopening or mask mandates or restricting group indoor activities like bars and dining.

One experiment to conduct and resolve this question is to look into the data at a county level. Different counties in the same state have had different levels of infection, and if state policies don't matter as much as herd immunity levels, then the spread of the disease in different counties post-Memorial Day would be very different, depending on the infection level. Moreover, different counties in the different clusters we identified can likely have similar behavior to each other, as long as the infection level was similar on Memorial Day.

\begin{figure}[]
\centering
\subfigure[Los Angeles-California vs Niagara-New York]{\includegraphics[width=0.9\linewidth, angle=0]{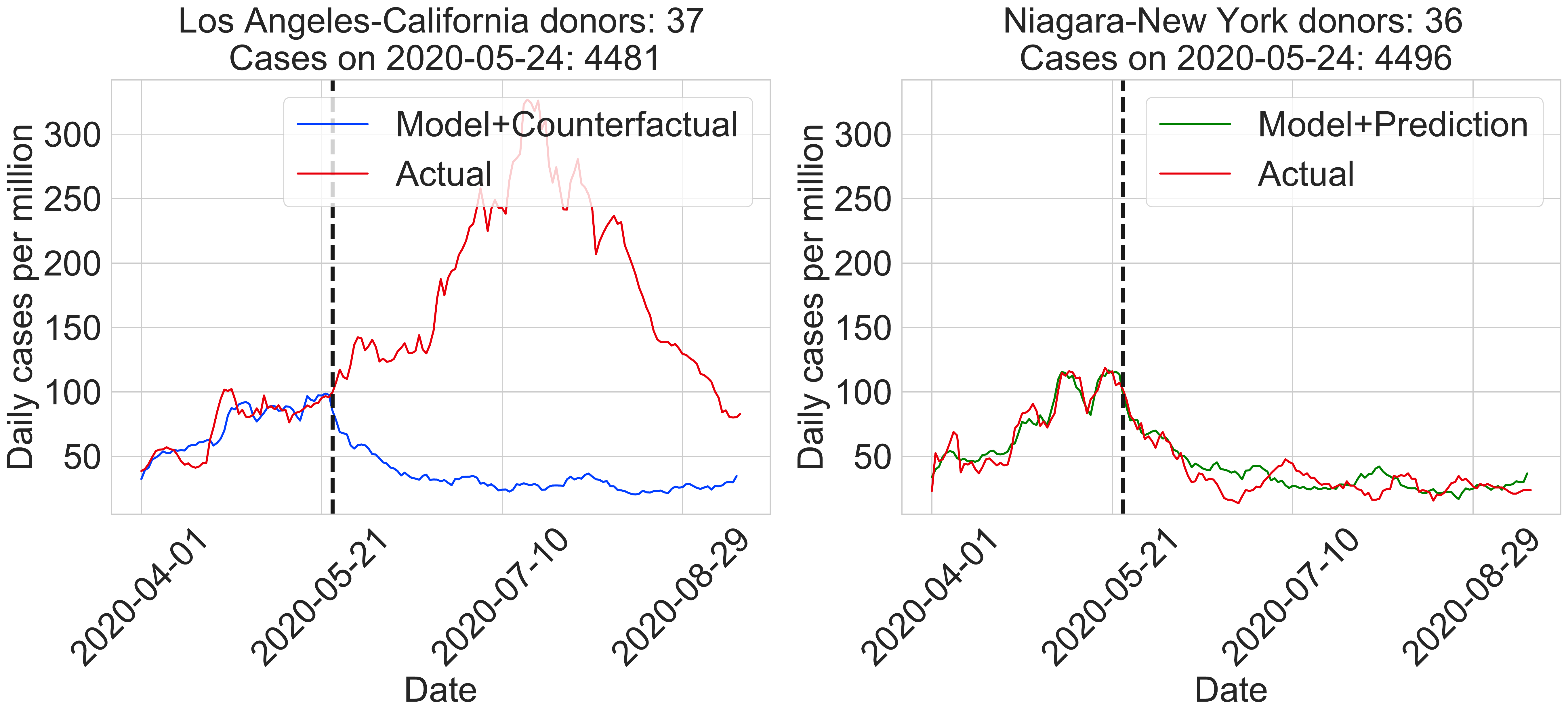}}

\subfigure[Broward-Florida vs Cumberland-Maine]{\includegraphics[width=0.9\linewidth, angle=0]{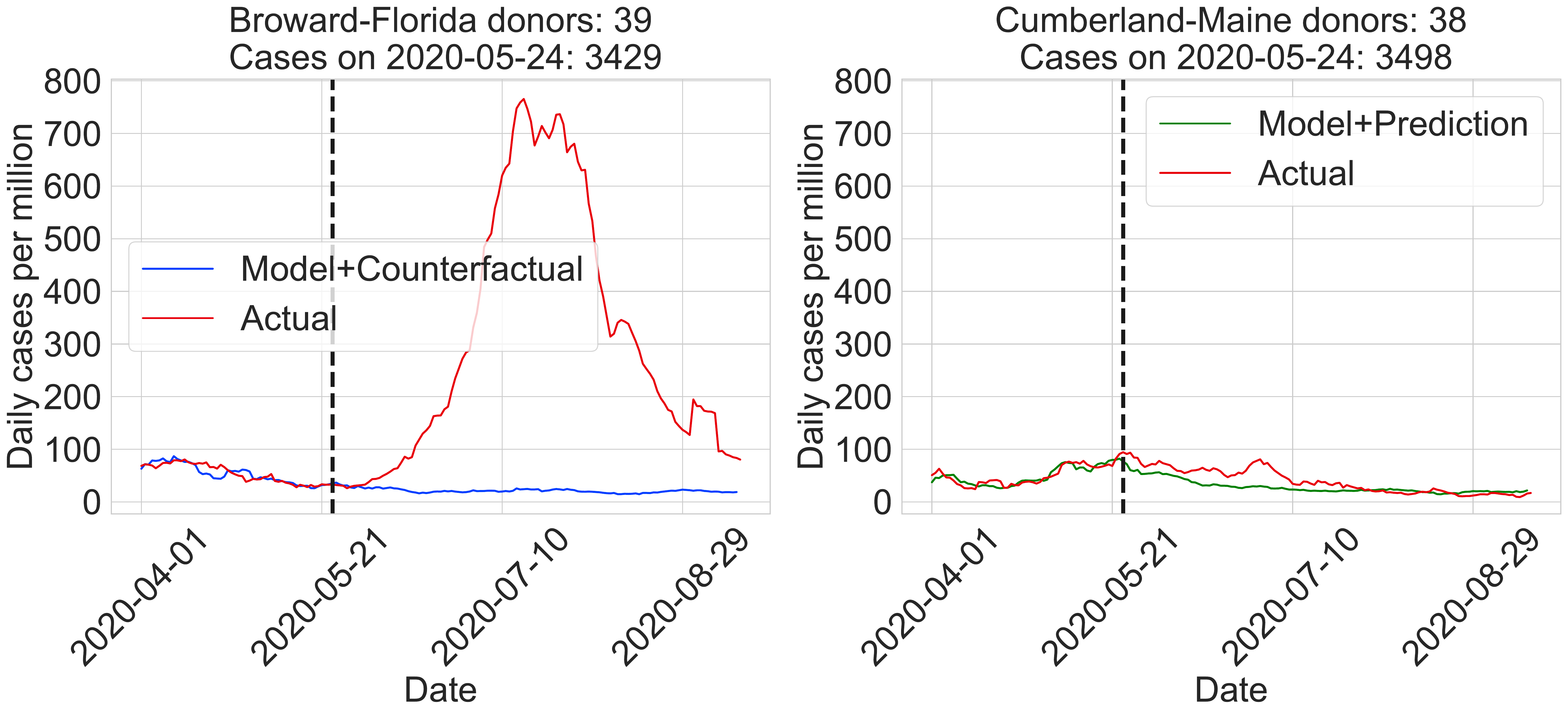}}

\subfigure[DeKalb-Georgia vs Madison-New York]{\includegraphics[width=0.9\linewidth, angle=0]{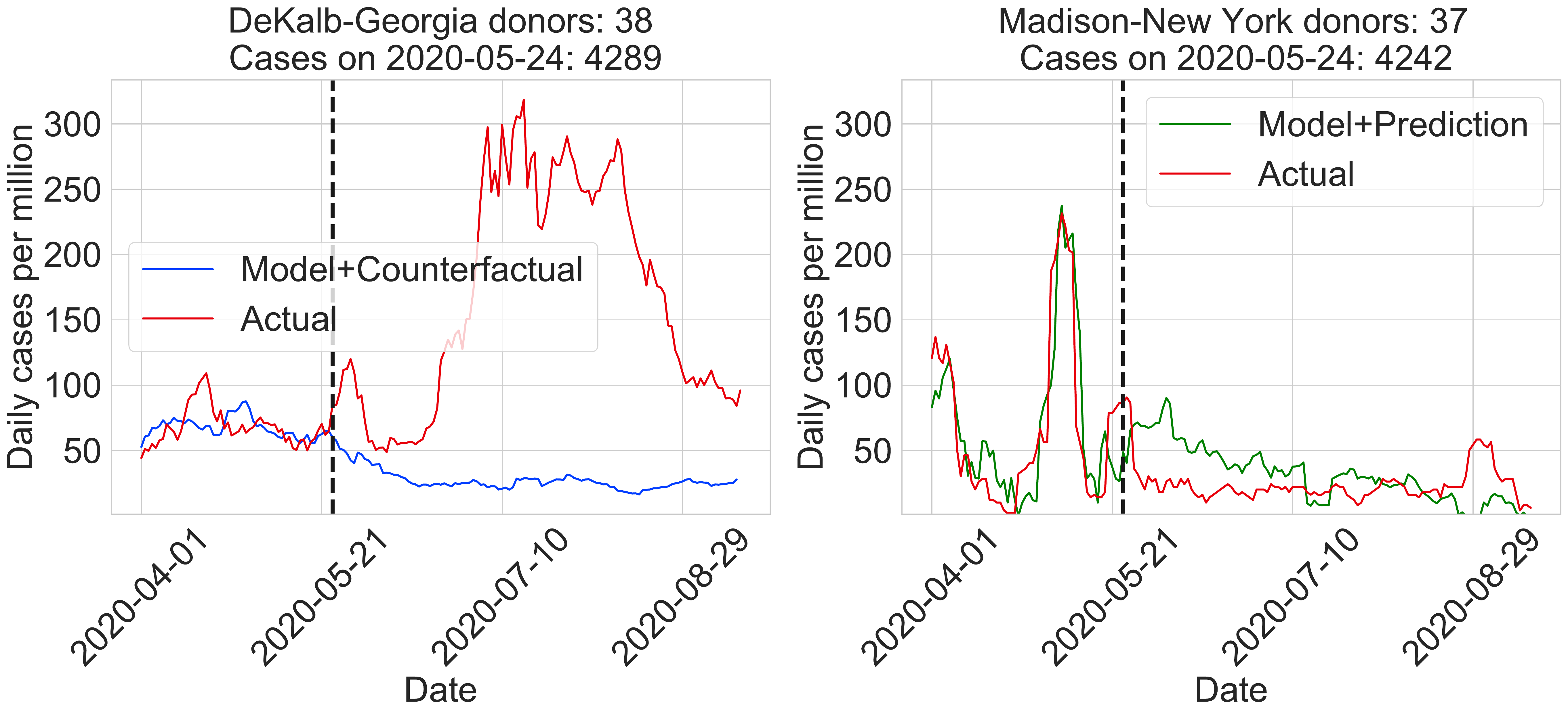}}

\subfigure[Bartholomew-Indiana vs Albany-New York]{\includegraphics[width=0.9\linewidth, angle=0]{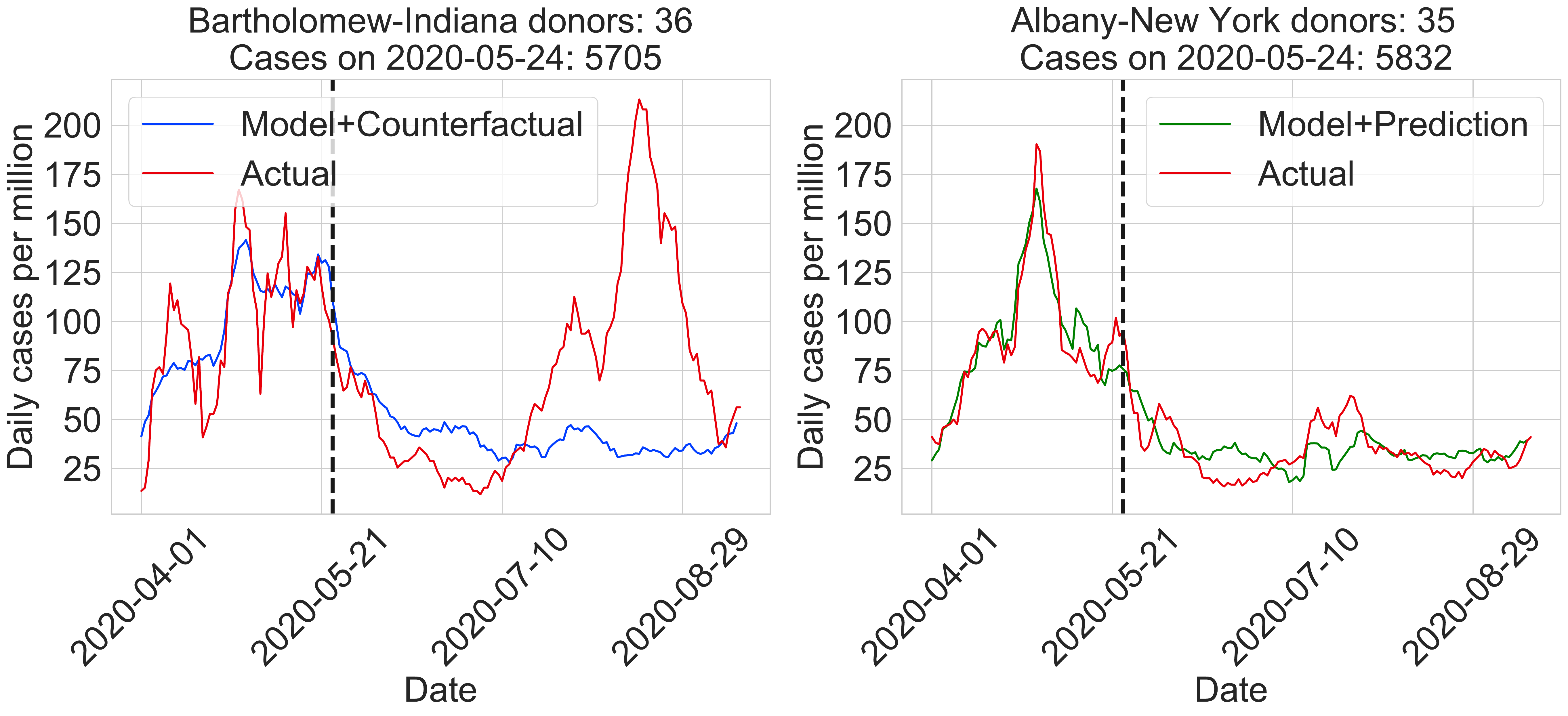}}

\subfigure[Jefferson-Louisiana vs Providence-Rhode Island]{\includegraphics[width=0.9\linewidth, angle=0]{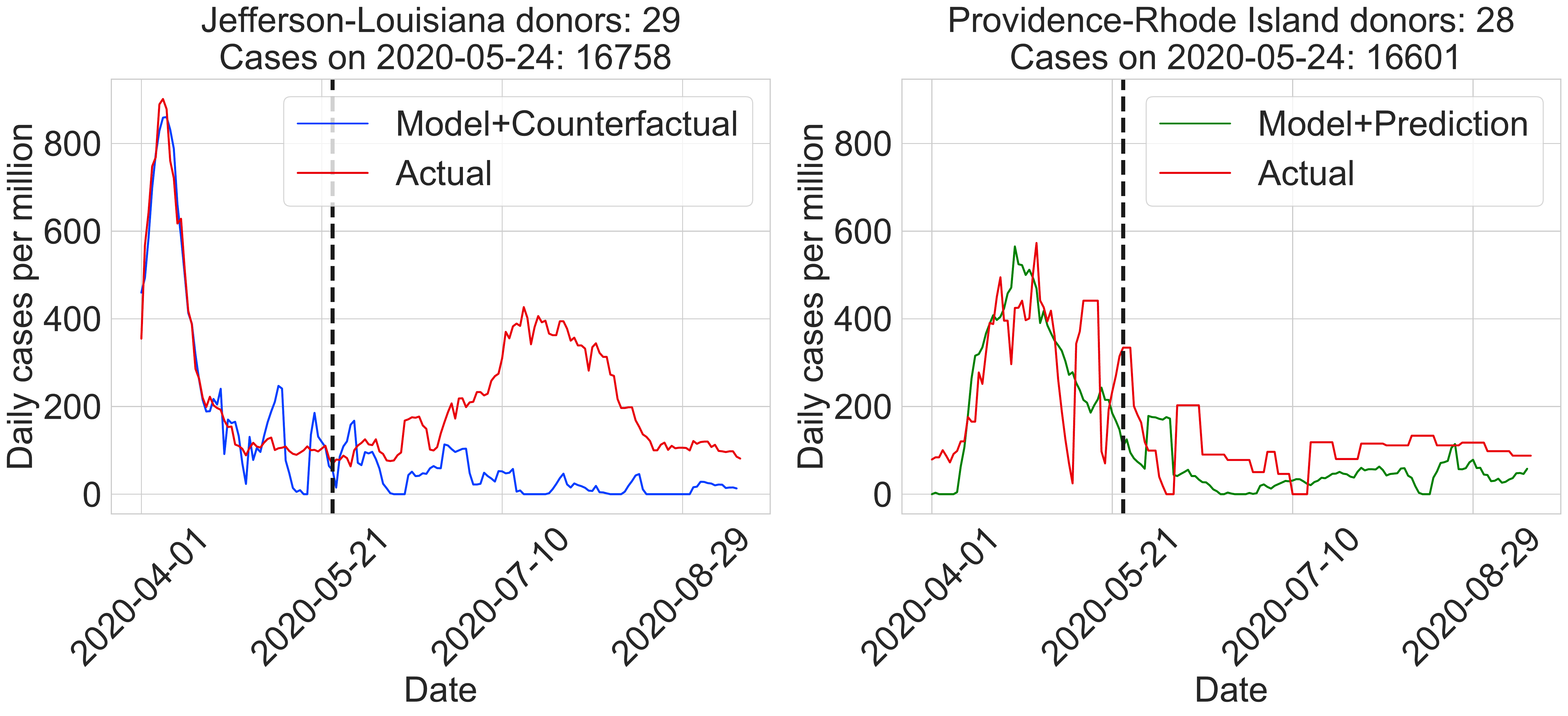}}
\caption{comparison of synthetic control counterfactual model of five counties within Northeast with five counties outside Northeast, with Northeast counties being the donor pool.}
\label{fig:herd_immunity}
\end{figure}

We first begin by identifying counties in the North-East where the virus hits early and compare them to similar counties in terms of infection cases in the rest of the US. Figure \ref{fig:herd_immunity} depicts different comparisons of this type. We picked counties with varying levels of cases per million on Memorial Day and constructed synthetic control models for all of them based on a donor pool of counties from the North-East cluster. This is in effect demonstrating a Synthetic Intervention \cite{SyntheticInterventions}, i.e. behavior of those counties had they mandated rules similar to the ones in the North-East. In each of the examples depicted, we can observe that the prediction for the counties in the North-East cluster closely follows the actual behavior, whereas, for all the example counties picked from other regions, there is a significant departure (for the worse) from the counterfactual of the actual cases. This indicates that the policies implemented in the North-East states \emph{did} have a significant effect on the spread of the virus, and the difference in behavior from the other counties cannot be explained by infection level or "herd immunity" alone.

To move from the anecdotal to statistical evidence of the impact of herd immunity, and whether or not, in reality, it impacts the spread we look at data across \emph{all} counties of the US. We examine if it plays a part, and if so at what level, and we again start with the counties that have a similar number of infections per million on Memorial Day, and pick a donor pool from the North-East. We compare the predictions for the North East counties vs counterfactual for the other counties, and compare the mean squared errors across the two sets. To see at what level herd immunity may occur, we use different bins of cases per million, e.g. $6000$, $8000$, $10000$, $12000$, etc. and see if at any number of cases, will the synthetic control model versus actual model look similar across different regions.

\begin{figure}[t]
\centering
{\includegraphics[width=0.7\linewidth, angle=0]{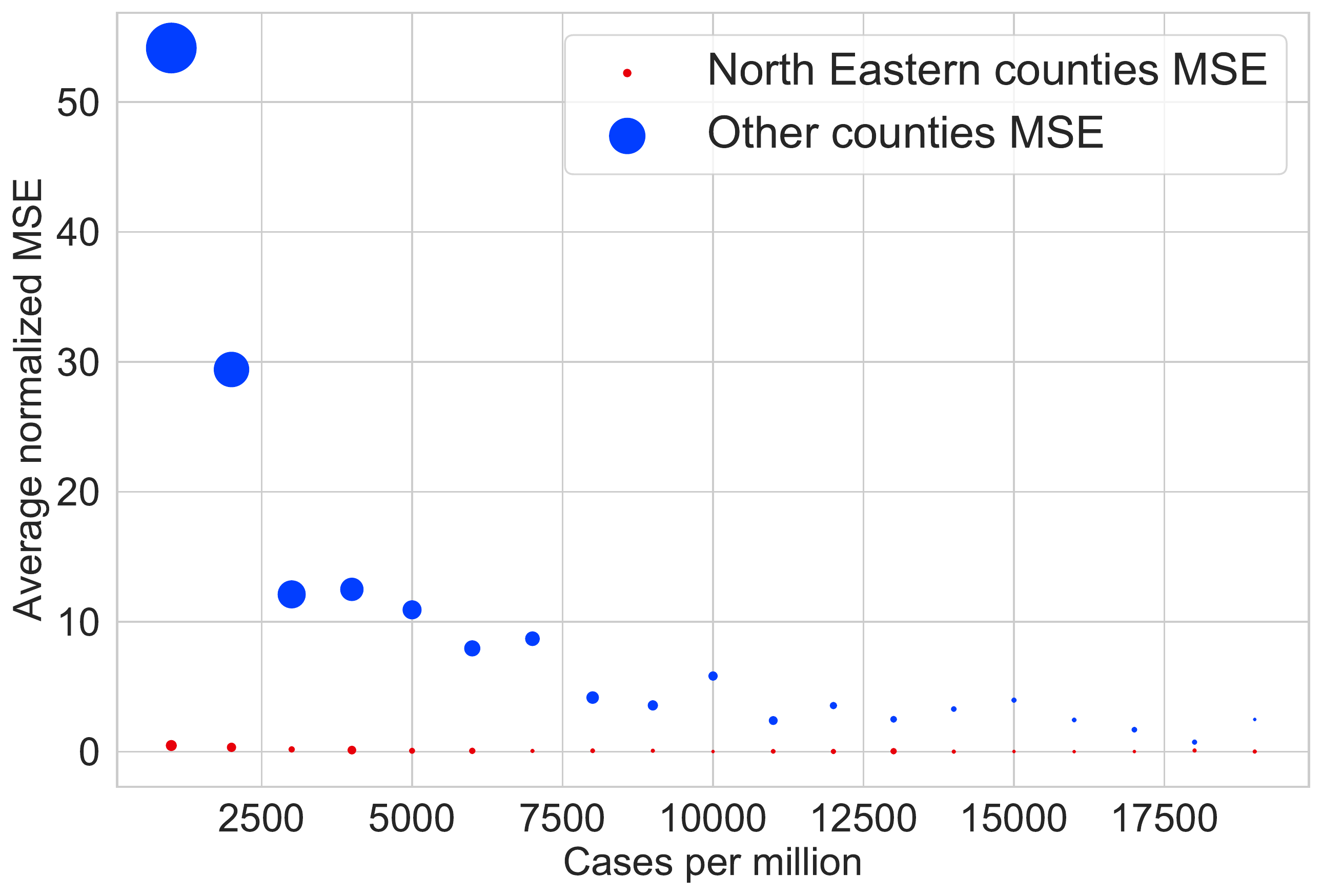}}
\caption{Average Mean squared error of the predictions for North-East and counter factual synthetic intervention for other counties, plotted as a function of case density on memorial day. The size of the marker reflects the number of counties represented in that data point}.
\label{fig:infections_compare}
\end{figure}

Figure \ref{fig:infections_compare} depicts a comparison of the normalized mean square error (normalized by total number of cases on Memorial Day) of the prediction versus actual cases for the North-East counties and for that of other counties. 
We observe that there is a large gap between the mean squared error of the predictions for the North-East and the other regions on the graph, which illustrates how strongly the social distancing policies can impact the spread given the different density of cases. This also confirms that the lockdown policies, including indoor dining prohibition and mask mandates, have a much stronger impact on the spread compared to "herd immunity" or infection levels. As the case density starts to increase, there is a general trend of reduction of the difference, which is consistent with the models (e.g., SIER) that predict a natural slowing down of the spread of the virus as the number of susceptible people decreases, however as we can see by the sizes of the sets of counties (represented as the size of the markers), the vast majority of the country is far away from any herd immunity effects.

\section{Conclusion}\label{sec:conclusion}
Synthetic Controls and Synthetic Interventions is a powerful technique to do counterfactual analysis and predictions. For COVID-19, they enable quick analysis of policy choices and predicts the number of cases/deaths under different circumstances. In our paper, we demonstrated the ability of the synthetic control method to predict and analyze the spread of COVID-19. Our models indicate that an earlier lockdown could have resulted in a significantly lower death count in New York, and this is true of most other places. Further, by analyzing the post-Memorial Day data at a county level, we show that intervention policies like mask mandates, mobility, and indoor dining/bar rules matter significantly in the spread of COVID-19. Counties that have stricter lockdown and social distancing mandates behaved significantly better than equivalent counties (i.e. with a similar infection level per capita) elsewhere with looser rules.

\section{Acknowledgements}
We thank Prof. Ambraham Flaxman of IHME for providing us with the lockdown dates dataset.

\bibliographystyle{unsrt}
\bibliography{main}

\begin{thebibliography}{10}

\bibitem{Mayooutbreak}
Mayo~Clinic Staff.
\newblock Coronavirus disease 2019 (covid-19).
\newblock
  \url{https://www.mayoclinic.org/diseases-conditions/coronavirus/symptoms-causes/syc-20479963}.

\bibitem{MafirstCOVID}
Josephine Ma.
\newblock Coronavirus: China’s first confirmed covid-19 case traced back to
  november 17.
\newblock
  \url{https://www.scmp.com/news/china/society/article/3074991/coronavirus-chinas-first-confirmed-covid-19-case-traced-back}.

\bibitem{CSSEJHU}
ArcGIS.
\newblock Covid-19 dashboard by the center for systems science and engineering
  (csse) at johns hopkins university (jhu).
\newblock
  \url{https://gisanddata.maps.arcgis.com/apps/opsdashboard/index.html#/bda7594740fd40299423467b48e9ecf6}.

\bibitem{US_regions}
Census regions and divisions of the united states.
\newblock
  \url{https://www2.census.gov/geo/pdfs/maps-data/maps/reference/us_regdiv.pdf}.

\bibitem{abadie2003economic}
Alberto Abadie and Javier Gardeazabal.
\newblock The economic costs of conflict: A case study of the basque country.
\newblock {\em American economic review}, 93(1):113--132, 2003.

\bibitem{abadie2010synthetic}
Alberto Abadie, Alexis Diamond, and Jens Hainmueller.
\newblock Synthetic control methods for comparative case studies: Estimating
  the effect of california’s tobacco control program.
\newblock {\em Journal of the American statistical Association},
  105(490):493--505, 2010.

\bibitem{young2014improving}
Scott~WH Young.
\newblock Improving library user experience with a/b testing: Principles and
  process.
\newblock {\em Weave: Journal of Library User Experience}, 1(1), 2014.

\bibitem{chaplin2006placebo}
Steve Chaplin.
\newblock The placebo response: an important part of treatment.
\newblock {\em Prescriber}, 17(5):16--22, 2006.

\bibitem{Amjad}
M.~Amjad, D.~Shah, and D.~Shen.
\newblock Robust synthetic control.
\newblock {\em J. Mach. Learn. Res.}, 19(1):802--852, January 2018.

\bibitem{SyntheticInterventions}
Anish Agarwal, Abdullah Alomar, Romain Cosson, Devavrat Shah, and Dennis Shen.
\newblock {Synthetic Interventions}.
\newblock \url{https://arxiv.org/pdf/2006.07691.pdf}.

\bibitem{donohue2019right}
John~J Donohue, Abhay Aneja, and Kyle~D Weber.
\newblock Right-to-carry laws and violent crime: A comprehensive assessment
  using panel data and a state-level synthetic control analysis.
\newblock {\em Journal of Empirical Legal Studies}, 16(2):198--247, 2019.

\bibitem{cunningham2018decriminalizing}
Scott Cunningham and Manisha Shah.
\newblock Decriminalizing indoor prostitution: Implications for sexual violence
  and public health.
\newblock {\em The Review of Economic Studies}, 85(3):1683--1715, 2018.

\bibitem{bohn2014did}
Sarah Bohn, Magnus Lofstrom, and Steven Raphael.
\newblock Did the 2007 legal arizona workers act reduce the state's
  unauthorized immigrant population?
\newblock {\em Review of Economics and Statistics}, 96(2):258--269, 2014.

\bibitem{kreif2016examination}
No{\'e}mi Kreif, Richard Grieve, Dominik Hangartner, Alex~James Turner, Silviya
  Nikolova, and Matt Sutton.
\newblock Examination of the synthetic control method for evaluating health
  policies with multiple treated units.
\newblock {\em Health economics}, 25(12):1514--1528, 2016.

\bibitem{heersink2017disasters}
Boris Heersink, Brenton~D Peterson, and Jeffery~A Jenkins.
\newblock Disasters and elections: Estimating the net effect of damage and
  relief in historical perspective.
\newblock {\em Political Analysis}, 25(2):260--268, 2017.

\bibitem{kermack27}
William~Ogilvy Kermack and Anderson~G McKendrick.
\newblock A contribution to the mathematical theory of epidemics.
\newblock {\em Proceedings of the royal society of london. Series A, Containing
  papers of a mathematical and physical character}, 115(772):700--721, 1927.

\bibitem{Prop99}
Stanton A~Glantz James M~Lightwood, Alexis~Dinno.
\newblock Effect of the california tobacco control program on personal health
  care expenditures.
\newblock
  \url{https://journals.plos.org/plosmedicine/article?id=10.1371/journal.pmed.0050178}.

\bibitem{usvt}
S.~Chatterjee.
\newblock Matrix estimation by universal singular value thresholding.
\newblock {\em Annals of Statistics}, 43:177--214, 2015.

\bibitem{LeeLiShahSong16}
C.~E. Lee, Y.~Li, D.~Shah, and D.~Song.
\newblock Blind regression: Nonparametric regression for latent variable models
  via collaborative filtering.
\newblock In {\em Advances in Neural Information Processing Systems 29}, pages
  2155--2163, 2016.

\bibitem{aldous}
D.~J. Aldous.
\newblock Representations for partially exchangeable arrays of random
  variables.
\newblock {\em Journal of Multivariate Analysis}, 11(4):581--598, 1981.

\bibitem{hoover1}
D.~N. Hoover.
\newblock Relations on probability spaces and arrays of random variables.
\newblock 1979.

\bibitem{hoover2}
D.~N. Hoover.
\newblock Row-columns exchangeability and a generalized model for
  exchangeability.
\newblock {\em Exchangeability in probability and statistics}, (281-291), 1981.

\bibitem{IHME}
Institute for Health~Metrics and Evaluation (IHME).
\newblock Covid-19 resources.
\newblock \url{http://www.healthdata.org/covid}.

\bibitem{Gonzalez-Reiche297}
Ana~S. Gonzalez-Reiche, Matthew~M. Hernandez, Mitchell~J. Sullivan, Brianne
  Ciferri, Hala Alshammary, Ajay Obla, Shelcie Fabre, Giulio Kleiner, Jose
  Polanco, Zenab Khan, Bremy Alburquerque, Adriana van~de Guchte, Jayeeta
  Dutta, Nancy Francoeur, Betsaida~Salom Melo, Irina Oussenko, Gintaras Deikus,
  Juan Soto, Shwetha~Hara Sridhar, Ying-Chih Wang, Kathryn Twyman, Andrew
  Kasarskis, Deena~R. Altman, Melissa Smith, Robert Sebra, Judith Aberg,
  Florian Krammer, Adolfo Garc{\'\i}a-Sastre, Marta Luksza, Gopi Patel, Alberto
  Paniz-Mondolfi, Melissa Gitman, Emilia~Mia Sordillo, Viviana Simon, and Harm
  van Bakel.
\newblock {}introductions and early spread of sars-cov-2 in the new york city
  area.

\bibitem{eaterjun2020}
Elazar Sontag.
\newblock Where restaurants have reopened across the u.s.
\newblock
  \url{https://www.eater.com/21264229/where-restaurants-reopened-across-the-u-s}.

\bibitem{herdimmunity}
David Dowdy and Gypsyamber D'Souza.
\newblock {What is Herd Immunity and How Can We Achieve It With COVID-19?}
\newblock
  \url{https://www.jhsph.edu/covid-19/articles/achieving-herd-immunity-with-covid19.html}.

\end{thebibliography}

\end{document}